\documentclass[twocolumn,twocolappendix]{aastex631}

\usepackage{amsmath, amssymb}
\usepackage{graphicx}
\usepackage{comment, url}
\usepackage{booktabs}
\usepackage{multirow}

\newcommand{\nh}{ n_{\rm H}}

\newcommand{\HI}{H{\sc ~i}}

\newcommand{\HeII}{He{\sc ~ii}}

\newcommand{\CII}{C{\sc ~ii}}
\newcommand{\CIII}{C{\sc ~iii}}
\newcommand{\CIV}{C{\sc ~iv}}

\newcommand{\NII}{N{\sc ~ii}}
\newcommand{\NIII}{N{\sc ~iii}}
\newcommand{\NV}{N{\sc ~v}}

\newcommand{\OI}{O{\sc ~i}}

\newcommand{\OVI}{O{\sc ~vi}}

\newcommand{\SiII}{Si{\sc ~ii}}
\newcommand{\SiIII}{Si{\sc ~iii}}
\newcommand{\SiIV}{Si{\sc ~iv}}
\newcommand{\SiV}{Si{\sc ~v}}
\newcommand{\SII}{S{\sc ~ii}}
\newcommand{\SIII}{S{\sc ~iii}}
\newcommand{\MgI}{Mg{\sc ~i}}
\newcommand{\MgII}{Mg{\sc ~ii}}
\newcommand{\MgX}{Mg{\sc ~x}}
\newcommand{\NeVIII}{Ne{\sc ~viii}}

\newcommand{\NH}{N_{\rm H}}
\newcommand{\fmm}{f_{\rm M}}

\newcommand{\cmv}{\rm cm^{-3}}
\newcommand{\cmc}{\rm cm^{-2}}
\newcommand{\kms}{\rm km\:s^{-1}}
\newcommand{\ergs}{\rm erg\:s^{-1}}
\newcommand{\msun}{\rm M_{\odot}}
\newcommand{\msuny}{\rm M_{\odot}\:yr^{-1}}
\newcommand{\tcool}{t_{\rm cool}}
\newcommand{\lcool}{L_{\rm cool}}
\newcommand{\mcool}{M_{\rm cool}}
\newcommand{\forg}{f_{\rm X,i}}
\newcommand{\feff}{\tilde{f}_{\rm X,i}}

\newcommand{\rvir}{R_{\rm vir}}
\newcommand{\teq}{T_{\rm eq}}
\newcommand{\thot}{T_{\rm hot}}

\newcommand{\vtu}{v_{\rm turb}}
\newcommand{\ttu}{t_{\rm turb}}
\newcommand{\ltu}{L_{\rm turb}}
\newcommand{\cs}{c_{\rm s}}
\newcommand{\mach}{{\cal{M}}}
\newcommand{\kb}{k_{\rm B}}

\shorttitle{Signatures of density fluctuations in CGM absorption}
\shortauthors{Faerman et~al.}

\begin{document}

\title{The signatures of density fluctuations and mixing gas in circumgalactic absorption systems}

\correspondingauthor{Yakov Faerman}
\email{yakov.faerman@gmail.com}
\author{Yakov Faerman}
\affiliation{University of Washington, Department of Astronomy, 3910 15th Ave NE, Seattle, WA, 98195, USA}
\author{Daniel R. Piacitelli}
\affiliation{Rutgers, The State University of New Jersey, Department of Physics and Astronomy, Piscataway, NJ 08854, USA}
\affiliation{University of Washington, Department of Astronomy, 3910 15th Ave NE, Seattle, WA, 98195, USA}
\author{Matthew McQuinn}
\affiliation{University of Washington, Department of Astronomy, 3910 15th Ave NE, Seattle, WA, 98195, USA}
\author{Jessica K. Werk}
\affiliation{University of Washington, Department of Astronomy, 3910 15th Ave NE, Seattle, WA, 98195, USA}

\defcitealias{Werk13}{W13}
\defcitealias{Werk14}{W14}
\defcitealias{Prochaska17}{P17}
\defcitealias{KS19}{KS19}
\defcitealias{FW23}{FW23}

\begin{abstract}
We investigate the prospects for detecting and constraining density and temperature inhomogeneities in the circumgalactic medium (CGM) using absorption measurements of metal ions. Distributions in the gas thermal properties could arise from turbulence, gas cooling from the hot phase, and mixing between the cool and hot phases. Focusing on these physically motivated models, we parameterize each with a single parameter for simplicity and provide empirical and theoretical estimates for reasonable parameter values. We then construct the probability distribution functions for each of these scenarios, calculate the effective ion fractions, and fit our models to the COS-Halos absorption measurements to infer the gas densities and metallicities. We find that the models we consider (i) produce similarly good fits to the observations with or without distributions in the gas thermal properties, and (ii) result in detectable changes in the column densities only at the boundaries of reasonable parameter values. We show that \HeII\ self-shielding can have a larger effect on the ion fractions than density and temperature fluctuations. As a result, uncertainties in cloud geometry and their spatial distribution, affecting the details of radiation transfer, may obscure the effect of inhomogeneities.
\end{abstract}



\section{Introduction}\label{sec:intro}

The circumgalactic medium (CGM) is the multi-phase gaseous medium surrounding galaxies, likely extending all the way to the dark-matter halo virial radius \citep{2017ARA&A..55..389T,Schaan21,Bregman22,Wilde23}. It serves as the channel through which matter can flow from the intergalactic medium, and where energy and metals from the galaxy are deposited. Thus, the CGM is critical in understanding galaxies and their evolution. Yet, exactly how the gas in the CGM is arranged, how it accretes onto galaxies, and what physical processes are at play are all questions that have challenged the CGM community (see \citealt{2017ARA&A..55..389T} and \citealt{2023ARA&A..61..131F} for recent reviews). One approach to addressing them is to pursue new observables, and an exciting set of observables are coming to the table, including mapping line emission \citep{2016ApJ...827..148C, 2022MNRAS.516.3049P, 2023arXiv231100856N}, Sunyaev-Zeldovich cosmic microwave background anisotropies \citep{Schaan21, Bregman22}, and Fast Radio Burst dispersion measures \citep{2023arXiv230101000R, 2023ApJ...945...87W}. Another approach is to buckle down and try to make sense of the observables we already have, with absorption measurements being the most abundant and sensitive to diffuse gas. This paper falls into this latter category.

The last decade has seen a large increase in the quality and quantity of CGM absorption data sets \citep[e.g.][]{Tumlinson11, Stocke13, Werk13, Bordoloi14, Borthakur15, Burchett19, Chen20, Zahedy21, Wilde21, KT23}, catalyzed by the installation of the UV-sensitive Cosmic Origins Spectrograph on the Hubble Space Telescope in 2009 \citep{2012ApJ...744...60G}. Observations of UV absorption are primarily sensitive to the cool, $T\approx 10^4$~K, phase of the CGM\footnote{Except for the \OVI\ absorption that likely probes the hotter gas (\citealp{FSM17,mcquinn18,Qu24}), and \NeVIII\ and \MgX\ absorption that is observed at (slightly) higher redshifts \citep{Meiring13,Muzahid13,Qu16}}. One major result from these studies is that this cool gas is common and abundant around star-forming and quiescent galaxies in galaxies with stellar masses similar to the Milky Way (MW) and estimated dark matter halo masses of $\sim 10^{12}~\msun$ (and see \citealp{Zheng24} for systems with lower mass halos). Furthermore, the fraction of the baryons residing in this phase may be large, with estimates ranging up to $100\%$ \citep{Werk14, Prochaska17}, although see \citealp{Stern16,FW23} for lower fractions). A second important result arising from the inferred low densities is that the cool gas may be out of thermal pressure equilibrium with the virialized gas (\citealp{Werk16, FW23}, but see \citealp{Qu23}). The latter inference has motivated many theoretical studies of the effect of nonthermal pressure support, especially owing to cosmic rays \citep{2018ApJ...868..108B, 2020MNRAS.496.4221J}, suggesting very different pictures for the CGM. Overall, our understanding of the cool CGM gas is far from complete, and we still do not have an understanding of how the cool gas is formed and survives, relates to other gas phases, or accretes onto the host galaxy (see \citealp{Fraternali17}).

Inferences about the cool gas properties use photoionization models, typically employing the omnibus \textsf{Cloudy} code \citep{Ferland17}. These models usually assume a slab density profile, with the incident radiation field perpendicular to the slab to model attenuation. They also assume that the gas is in photoionization equilibrium (PIE) and its temperature is set by cooling balancing the heating from the UV background. In this regime the equilibrium temperature is set by only the gas density and metallicity (although some studies leave the temperature as a free parameter, e.g. \citealp{Zahedy21, Haislmaier21}). The model is then fit to the measurements, either of the total columns of hydrogen and metal ions \citep[e.g.][]{Werk14,Prochaska17}, effectively constraining the average gas properties along the line of sight (see \citealp{Hafen24}), or of individual kinematic components \citep[e.g.][]{Zahedy21,Qu23}, possibly corresponding to clouds or cloud complexes (although see \citealt{Peeples19}).

There has been recent work to both generalize the model assumptions, as well as to test whether these assumptions hold in simulations. For example, \citealp{Stern16}) assume a wide distribution of densities to account for the different ionization states observed. \citet{Haislmaier21} fit a multi-component model to the total columns observed in CGM absorption systems and found that all their fits preferred 2-3 phases. \citet[hereafter FW23]{FW23} constructed a radial density profile for cool gas embedded in the hot ambient CGM and forward-model the ion column densities to estimate the range of CGM properties such as spatial extent, density, and mass. \citet{Hafen24} modeled mock absorption spectra generated by toy models and detailed CGM simulations and found that the fits generally retrieved the correct properties (densities, temperatures, and metallicities) of the dominant components, but were less successful detecting the more minor contributions of smaller clouds or more diffuse gas.

In this paper we investigate how different scenarios that lead to distributions in gas density and temperature
may manifest in the ionic columns observed in absorption from cool CGM gas. The manuscript is organized as follows: Section~\ref{sec:scenarios} discusses physical scenarios that may shape the distribution of cool and intermediate temperature gas. Motivated by this discussion, Section~\ref{sec:methods} then devises physically-motivated parameterizations for the gas distributions and presents calculations for how the effective ion fractions in these models contrast with those in single temperature and density models. In Section~\ref{sec:fit_testing}, we investigate whether the differences with single-density models are observable or possibly even favored by data, by fitting these models to measurements of ionic columns from the COS-Halos survey. We address the effects of metallicity, equilibrium temperature, and potential self-shielding by \HeII\ on our results. In Section~\ref{sec:discussion}, we discuss the scaling of our results with redshift, compare them to results of previous work, and make predictions for possible follow-up observations. We summarize our findings in Section~\ref{sec:summary}.

\section{Scenarios for density and temperature distributions}\label{sec:scenarios}

 We now outline potential scenarios in the CGM which can lead to a distribution in both density and temperature. \S\ref{subsec:scen_turbulence} discusses lognormal density distributions, which may arise in isothermal and adiabatic turbulence, or as a superposition of many physical processes influencing the CGM. \S\ref{subsec:scen_int} considers density and temperature distributions motivated by cooling gas and mixing layers.  This discussion motivates the parameterizations that we will later adopt -- as well as plausible ranges for the parameters -- to calculate the effect of such distributions on the abundances of different ions.
 
Our discussion will often reference the cooling time and the length over which gas is able to maintain acoustic contact during cooling and so we define them here:
\begin{equation}
\tcool = \frac{5}{2} \frac{n}{n_e}\frac{k_B T}{\nh \Lambda} = 2.0 ~{\rm Myr}~ \left( T_{4} {n_{-3}^{-1} \Lambda_{-22}^{-1}}\right), ~~~
\label{eqn:tcool}
\end{equation}
\begin{equation}
\lcool \equiv \cs \tcool = \frac{5}{2} \frac{n}{n_e}\frac{\bar{m}\cs^3}{\gamma \nh \Lambda} = 32 ~{\rm pc}~ \left( T_{4}^{3/2} {n_{-3}^{-1} \Lambda_{-22}^{-1}}\right),
\label{eqn:lcool}
\end{equation}
where $T_4$ is the temperature in units of $10^4$~K, approximately the heating/cooling equilibrium temperature in photoionized gas (see Figure~\ref{fig:temp_eq} and \S\ref{sec:methods}), $\Lambda_{-22}$ is the cooling rate coefficient in units of $ 10^{-22}$~erg~$\cmv$, roughly its value for $T=10^{4}$~K gas, $n_{-3}$ is the (total) density of hydrogen in units of $10^{-3}~\cmv$, characteristic of the inferred density of low-z CGM absorption systems which have $10^{-4}-10^{-2}~\cmv$ \citep{Werk14,Prochaska17}, $\bar{m} = 0.59 m_{\rm H}$ is the mean particle mass for fully ionized gas, and $\cs$ is the gas sound speed, given by $\cs = \sqrt{\gamma k_BT/\bar{m}} = 15.2~{\kms}~T_4^{1/2}$ for $\gamma=5/3$\footnote{These $\tcool$ and $\lcool$ are for isobaric gas, relevant for our discussion later on; for isochoric cooling they would be $3/5$ smaller.}.

\subsection{Lognormal models and turbulence}\label{subsec:scen_turbulence}

Perhaps the most generic distribution is a lognormal probability distribution function (PDF) in the density.
This PDF, for example, is found for both the cool and hot phases of the CGM in the Illustris-TNG galaxy formation simulation \citep{Dutta24}. While such a PDF may result from the complexity of the CGM and the central limit theorem, one physical motivation for a lognormal distribution is turbulence \citep[e.g.][]{2010A&A...512A..81F}.  In what follows, we outline when CGM turbulence is likely to result in isothermal or adiabatic turbulence and generate a lognormal distribution with the standard deviation in density given by $\sigma$.

First, if the cooling time is shorter than the timescale for turbulent eddies to turnover, the turbulence will be nearly isothermal\footnote{We state `nearly isothermal' for the $\sim 10^4$~K phase of relevance to this study because the gas in this slow eddy limit will tend to the equilibrium temperature, $\teq$, where photoheating balances cooling, which at the low densities relevant for the CGM does have a mild density trend. As we show in the next section, the temperature is a weak function of the density (see Figure~\ref{fig:temp_eq}), and in our discussion here we will refer to this case as `isothermal' for simplicity of terminology while using $\teq$ for the gas temperature.}. Namely, the turbulence is approximately isothermal if the driving scale satisfies $\ttu \equiv \ltu/\vtu > \tcool$. Inserting $\vtu = \mach \cs$ we can write this condition as
\begin{equation}
\ltu \gtrsim 32 ~{\rm pc}~ \left(\frac{\mach T_{4}^{3/2}}{n_{-3}\Lambda_{-22}}\right) \approx 53 ~{\rm pc}~ \left(\frac{T_{4}v_{25}}{n_{-3}\Lambda_{-22}}\right) ~~~,
\label{eqn:ldrive}
\end{equation}
where we have again scaled quantities to values characteristic of the low-$z$ CGM absorption systems.

What Mach numbers are expected in the cool CGM? The line widths of metal absorption lines in the COS-Halos sample, probing the CGMs of low-redshift galaxies, are in the range of $10-45~\kms$ for many different ions, including \CII-\CIV, \SiII-\SiIV, and \NII-\NIII\ \citep[e.g.][]{Werk13,Qu22}\footnote{It is also interesting to note that these line widths are lower than those of the \OVI\ absorption lines, with a range of $10-100~\kms$, and a mean of $\left< b \right>\approx 50~\kms$, strong evidence that this ion is tracing hotter gas.}. It is unclear whether these widths are caused by a (supersonic) velocity dispersion in a single cloud, or motions of many physically unrelated clouds along the line of sight, each with a subsonic velocity dispersion. If the latter scenario is a better description of the real CGM, then our estimate is an upper limit for the Mach number. Motivated by these measurements in the second transition of Equation~\eqref{eqn:ldrive}, we scaled the equation above to a turbulent velocity of $25~\kms$, the mean Doppler parameter of the different ions in the COS-Halos sample. For the $T \approx 1-2\times 10^4$~K PIE gas that is likely to be responsible for the absorption, $\vtu = 25~\kms$ corresponds to $\mach \approx 1.2 - 1.7$ and the condition on the driving scale for isothermal turbulence is $\ltu \gtrsim 50-100$~pc.

Another check that the isothermal limit holds compares the dissipation rate of turbulent energy to the radiative cooling rate, which for gas at PIE is balanced by heating from the metagalactic radiation field (MGRF). If the dissipative heating is lower than the cooling rate, the gas temperature should be close to $\teq$ and hence be in our isothermal limit. We have compared these rates and find that, for the observationally-favored limit of $\mach \lesssim 2$, Equation~\eqref{eqn:ldrive} provides a stronger constraint on $\ltu$ than the constraint from energy considerations. However, for higher Mach numbers, energetics sets the eddy size that can stay isothermal, with it scaling instead as $\propto \mach^3$.

What turbulence driving scales do we expect to find in the CGM? A natural scale is the size of a cloud, as each cloud is being buffeted by winds and sheared by Kelvin-Helmholtz instabilities as it moves through the hot virialized gas. Estimates for the absorption path length from the total ionic columns across $z\sim 0.2$ CGM systems suggest maximum cloud sizes of  $1-100~$kpc \citep{Werk16, mcquinn18}. This is larger than the limit we estimate, suggesting CGM turbulence is isothermal. However, many clouds could contribute to the total columns. To satisfy the limit in Equation~\eqref{eqn:ldrive}, the number of clouds along a $\sim 10$~kpc path length should be below $\sim 100$ and, then, mildly supersonic turbulence can be isothermal. In the shattering clouds picture (the ``misty" CGM, see \citealp{mccourt18}), cloud sizes are approximately $\lcool$, which for cool CGM densities, $\nh \sim 10^{-4}-10^{-2}~\cmv$, translates to $\approx 5-500$~pc. This is identical to our limit for $\ltu$ for $\mach \sim 1$.

Next, we relate the turbulence Mach number to the width of the lognormal density distribution.

\subsubsection{Isothermal fluctuations}\label{subsec:fluc_isothermal}

Hydrodynamic studies show that isothermal turbulence is well characterized by a lognormal PDF with the variance in the natural logarithm of $\rho/\bar{\rho}$ given by
\begin{equation}\label{eqn:sigmach}
\sigma^2 = \ln (1 + b^2 \mach^2),
\end{equation}
where $b\approx 1$ for compressive driving of the turbulence and $b\approx 0.3$ for solenoidal driving \citep{1994ApJ...423..681V, Kritsuk_2007, 2010A&A...512A..81F}. Real driving is likely a mixture of these two modes and hence has an intermediate $b$. As we estimated earlier for $\vtu = 25~\kms$, $\mach = 1.2-1.7$ and for compressive driving this results in $\sigma \approx 0.9-1.2$.

As mentioned in the introduction, there is some evidence that cold clouds could have nonthermal pressure support, such as from magnetic fields. For low Mach numbers, the $\sigma-\mach$ relation in Equation~\eqref{eqn:sigmach} can be approximated as $\sigma \approx b \mach$. When magnetic fields are included, the low Mach number scaling changes to $\sigma \propto {\cal M}^2$ for the Mach numbers below the Alfv\'enic Mach number, and $\sigma$ latches onto its hydrodynamic form at higher ${\mach}$ \citep{2007ApJ...658..423K}, so our previous estimate is still valid. The PDF of turbulence would only be substantially changed for our estimate of $\mach \sim 1$ if the magnetic pressure is substantially higher than thermal.

\subsubsection{Adiabatic fluctuations}\label{subsec:fluc_adiabatic}

Driving at scales smaller than the limit given by Equation~\eqref{eqn:ldrive} would lead to compressions over a timescale that is shorter than the cooling time. Then the largest eddies, which are most important for setting the compressions, would behave adiabatically. As discussed above for densities characteristic of the low-$z$ CGM, this could occur if the cloud sizes are $\lesssim 50-100~$parsec for mildly supersonic turbulence driven on the cloud scale.

While the density PDF in the adiabatic case is still well characterized by a lognormal distribution, the density fluctuations (and hence $\sigma$) are smaller than in the isothermal case. \citet{Nolan15} find that density fluctuations for adiabatic turbulence in monatomic gas ($\gamma = 5/3$) can be well approximated by $ \sigma^2 = \ln (1 + b^2 \mach^{3})$ when $b\mach \lesssim 1$, where $b\sim 0.3$ for solenoidal. As a result, at $\mach \approx 1.2-1.7$, $\sigma \approx 0.4-0.6$. At very high Mach numbers, for which $b\mach \gtrsim 1$, they find that $\sigma$ saturates at $\sigma^2 \approx 1.4$ for solenoidal driving and likely at a somewhat higher value for compressive driving.

\subsubsection{Lognormal distributions in simulations}
\label{subsec:simulations}

Cosmological simulations of galaxy formation capture a wide range of phenomena affecting the distributions of CGM properties, including different turbulence-driving mechanisms such as IGM accretion, stellar and AGN-driven outflows, galaxy mergers, etc. (see \citealp{Fielding20sim} for a comparison of different simulations). For example, \citet{Ramesh23} focus on the properties of cool gas clouds around MW-mass galaxies in the TNG50 simulation and find density distributions with a full-width-half-maximum (FWHM) of $\approx 1$~dex, corresponding to $\sigma \approx 1$ in our parameterization (see their Figure~5). \citet{Dutta24} show the density and temperature distribution of the CGM for a single halo and find a similar width for the cool gas, $\sim 1$~dex (Figure~9). However, these cool gas properties in these simulations are almost certainly not converged with resolution \citep{Hummels19, Peeples19,Voort19,Nelson20, Ramesh24}.

\vspace{0.5cm}

In summary, the above motivates lognormal distributions with $0 \le\sigma \le 1.2$. It also suggests for the largest eddies that isothermal turbulence is more physically likely in the CGM than the adiabatic turbulence since the turbulent driving scale in the latter must be very small.

\subsection{Fluctuations driven by cooling/mixing}\label{subsec:scen_int}

So far we have concentrated on models in which the gas is at $\teq \sim 10^4$~K, motivated by the fact that gas tends to cool quickly to its equilibrium temperature. Furthermore, at these temperatures a significant fraction of the metals are in the low to intermediate ionization states that the UV observations primarily probe. However, it is possible that some of the column in intermediate ions, such as \SiIV, \CIV, and \NV, forms in intermediate temperature gas, residing in turbulent mixing layers around cool gas clouds or in gas that is cooling radiatively from the hot/virial phase. We now consider these models for intermediate temperature gas.

\subsubsection{Mixing layers}\label{subsec:int_mixing}

Turbulent and conductive mixing layers sit at the boundary of cool clouds as the gas transitions from the halo virial temperature to the temperature of the cloud \citep{1990MNRAS.244P..26B,  Fielding20,2023MNRAS.520.2148Y, 2023ARA&A..61..131F}. Hydrodynamic simulations find this transition is close to isobaric \citep{2019MNRAS.487..737J}, and the boundary layer gas could either be condensing onto the clouds or evaporating  \citep{2018MNRAS.480L.111G}. \cite{Tan21b} studied a range of physical parameters and found that these layers generically produce a nearly flat distribution in temperature by volume (see their Figure~4).

To understand the effects of mixing gas on the ion fraction and the observable column density, we add a boundary layer-motivated flat temperature distribution between $\teq$ and $\thot$ on top of a cool component. We adopt $\thot = 10^6$~K, roughly the virial temperature of MW-mass halos, but our results are only logarithmically sensitive to this choice (see \S\ref{sec:methods} and Appendix~\ref{app_dist}). The cool gas component drives most of the absorption, and we model it with a $\delta$-function at $\teq$. We then parameterize the amount of gas at intermediate temperatures by the mass-weighted ratio of mixing gas to cool gas, $\fmm$. This way we hold the amount of gas in the cool phase fixed and our calculations show how the addition of mixing gas affects the total column.

We now estimate the mass fraction in intermediate temperature gas. Using hydrodynamic simulations, \citet{2019MNRAS.487..737J} find that the width of the turbulent boundary layer can be approximated by
\begin{equation}\label{eqn:lmix}
\ell \approx 170 ~{\rm pc}~ [\Lambda_{-21.7}^{T_{\rm mixing}}]^{-1/2} n_{-4}^{-1/2} (T_{\rm mixing}/10^5 {\rm K})^{-1/2} ~~~, 
\end{equation}
where $T_{\rm mixing}$ is the geometric mean of the cool and virialized gas temperatures (such that $T_{\rm mixing}\sim 10^5$~K for Milky Way-like halos), $\Lambda_{-21.7}^{T_{\rm mix}}$ the cooling rate evaluated at $T_{\rm mixing} \sim 10^5$~K, and $\ell$ is the length-scale that spans the $10^4~{\rm K}-T_{\rm mixing}$ transition in the boundary layer. To estimate the cloud to layer mass ratio, we assume spherical clouds and layers, and then the planar geometry assumed by \citet{2019MNRAS.487..737J} implies $l/R_{\rm cl} \equiv x \ll 1$. The mass ratio in the mixing layer to cool gas can then be written as
\begin{equation}
\fmm \equiv \frac{M_{\rm mixing}}{M_{\rm cool}} = \frac{n_{\rm mixing}}{n_{\rm eq}} \left[(x+1)^3-1\right] \approx 3 x \frac{n_{\rm mixing}}{n_{\rm eq}} ~~~,
\end{equation}
where $n_{\rm mixing}$ and $n_{\rm eq}$ are the densities of mixing gas and the cool gas in PIE, and the last transition is an approximation for $x \ll 1$, consistent with \citet{2019MNRAS.487..737J}. We place a conservative upper limit on $\fmm$ by taking $l = R_{\rm cl}$, this yields $\fmm = 0.7$ for a density contrast of $n_{\rm mixing}/n_{\rm eq} = \teq/T_{\rm mixing} = 0.1$.

\subsubsection{Cooling gas}\label{subsec:int_cooling}

Another scenario for the CGM is that the hot gas cools radiatively and creates a distribution of intermediate temperatures with probabilities that are proportional to the local gas cooling time \citep{Heckman02, 2017ApJ...848..122B, mcquinn18,Qu18a}. We examine this scenario and in Appendix~\ref{app_dist} we show that for isobaric cooling with a mass cooling rate of $\dot M$, the mass in a logarithmic integral in temperature is given by $dM/d \ln T = \dot M \, \tcool$. \citet{mcquinn18} estimate that if cooling gas is the origin of the observed \OVI\ columns around Milky Way mass systems, $\dot M \sim 10-100~\msuny$ are required\footnote{The CGM mass cooling rate is not necessarily (and most probably not) equal to the star formation rate in the galaxy. The difference can be due to heating from supernova feedback \citep{mcquinn18, Pandya20, F22}, AGN feedback, or other sources, offsetting some of the radiative cooling. The net cooling rate can also be higher than the SFR as some of the gas that cooled can be disrupted on its way to the disk and mixed back into the ambient phase \citep{Joung12}. The measured star formation rates in the COS-Halos galaxies are $\lesssim 10~\msuny$ \citep{Werk12} and $\sim 2~\msuny$ in the MW \citep{Lic15,BHG16}.}. We can relate this to mass of the cool CGM gas, which studies infer to be $M_{\rm cool} \sim 10^{10}~ M_\odot$ around Milky Way-mass systems \citep{Werk14, Prochaska17, FW23}, by taking their ratio:
\begin{equation}\label{eqn:cool_est1}
\begin{split}
f_{d \ln T} &~ \equiv \frac{dM/d \ln T}{M_{\rm cool}} = \\
&~ = 0.11 \; T_{5} n_{-4}^{-1} \Lambda_{-21.7}^{-1} \left(\frac{\dot M}{10~\msuny} \right) \left(\frac{ M_{\rm cool}}{10^{10}~M_\odot} \right)^{-1} ~~~.
\end{split}
\end{equation}
In this estimate $f_{d \ln T}$ is the mass in a logarithmic temperature bin (the fraction of gas per $\log(e)$ dex). The actual amount of cooling gas may be larger, and the value of $f_{d \ln T}$ may be higher by a factor of a few. On the other hand, if the \OVI~traces a different gas phase, that is not actively cooling, this would lower $f_{d \ln T}$. Finally, $f_{d \ln T}$ relates the mass ratio in cooling to cool gas for an average sightline and specific sightlines may have a different ratio.

Another way to empirically estimate the amount of gas in the cooling gas is through the observed column densities. The column density of ion $i$ of an element $X$ can be written as $N_{\rm X,i} \propto M_{\rm gas} A_{\rm X} Z \forg$, where $M_{\rm gas}$ is the gas mass in a given phase, $A_{\rm X}$ is the elemental abundance (number density in $X$ relative to hydrogen), $Z$ is the gas metallicity relative to solar, and $\forg$ is the ion fraction. We can then write the \OVI\ to \NIII\ column density ratio, for example, as
\begin{equation}\label{eqn:cool_est2a}
\frac{N_{\rm \text{\OVI}}}{N_{\rm \text{\NIII}}} = \frac{M_{\rm cooling}}{M_{\rm cool}} \frac{A_{\rm O}}{A_{\rm N}} \frac{f_{\rm \text{\OVI}}}{f_{\rm \text{\NIII}}} \frac{Z_{\rm cooling}}{Z_{\rm cool}}.
\end{equation}
Inserting the solar ratio of nitrogen to oxygen abundances ($A_{\rm N}/A_{\rm O} \approx 1/7$, \citealp{Asplund09}), the mass ratio is then:
\begin{equation}\label{eqn:cool_est2}
\fmm \equiv \frac{M_{\rm cooling}}{M_{\rm cool}} = 0.14 \; \left(\frac{N_{\rm \text{\OVI}}}{N_{\rm \text{\NIII}}}\right) \left(\frac{f_{\rm \text{\OVI}}}{f_{\rm \text{\NIII}}}\right)^{-1} \left(\frac{Z_{\rm cooling}}{Z_{\rm cool}} \right)^{-1} ~~~.
\end{equation}
In the COS-Halos sample, \OVI~column density is a factor of $\sim 2$ higher than that of \NIII\ and ionization models suggest that the ratio of ion fractions is ${f_{\rm OVI}}/{f_{\rm NIII}} \sim 1$ \citep{mcquinn18}. Assuming similar metallicities in the cooling and cool gas gives $\fmm \approx 0.3$.

In this work we use $\fmm$ as the model parameter for cooling gas, where we only count gas up to \OVI-peak-abundance temperature of $T=2\times10^5$K as being in the cooling phase to better connect with the estimate given by equation~\eqref{eqn:cool_est1}. Since for metal-enriched gas this is the peak of the cooling curve, and most of the mass in the distribution resides at higher temperatures, this definition results in $f_M \approx f_{dlnT}$ -- i.e. the mass fraction in cool gas is similar to the amount of cooling gas in a logarithmic bin around  $T=2\times10^5$K, connecting this definition also with the estimate given by equation~(\ref{eqn:cool_est2}). We discuss the details of the cooling gas model PDF in \S\ref{subsec:distributions} and Appendix \ref{app_dist}.

We note that \OVI\ may form in other CGM phases, tracing gas that is not actively cooling \citep{Stern16}, or in which cooling is balanced by heating \citep{FSM20}. Ideally, we could use additional ions, such as \CIV~and \NV~for these estimates. However, measurements of \CIV~in the low redshift CGM are currently rare (although see \citealp{Berg_civil} for upcoming data) and \NV~observations provide mostly upper limits (see \citealp{Werk13}). Thus, we now use the measured OVI columns to place upper limits on the amount of cooling gas, and in this work we do not fit for the observed columns (see also \S\ref{sec:discussion}).

\vspace{0.5cm}
In conclusion, both estimates suggest that values of tenths for $\fmm$ are likely, and values as large as unity may be possible. We note that while having a large amount of mixing or cooling gas is energetically challenging, \citet{mcquinn18} find that for Milky Way-like galaxies there may be enough energy in stellar feedback to support a column of cooling gas that is comparable to the observed amount in cool gas. For example, star formation rates (SFR) of $1-10~\msuny$, typical of the COS-Halos galaxies, imply SNe rates of $0.01-0.1~{\rm yr^{-1}}$, with energy outputs of $\sim 10^{49}-10^{50}~{\rm erg~yr^{-1}}$. If $\approx 10\%$ of this energy input is coupled to the CGM, it provides a heating rate of $L_{\rm heat} \approx 3 \times 10^{40}-3 \times 10^{41} ~\ergs$. This can offset most of the radiative cooling of gas with $N_{\rm OVI} \sim 2 \times 10^{14}~\cmc$ in one of two scenarios (i) volume-filling gas in a $R \approx 150$~kpc CGM, or (ii) mixing layers with width $L/R_{\rm cl} = 0.3$ around cool gas clouds occupying $30\%$ percent of the same volume.

\section{Distributions and Ion fractions}\label{sec:methods}

We now describe our calculations of the ion column densities in the presence of density and temperature distributions. First, we construct the probability distributions for the scenarios described in \S\ref{sec:scenarios}, then, we calculate the effective ion fractions for these distributions. Our results are presented in Figures \ref{fig:eff_frac1} and \ref{fig:eff_frac2}.

\subsection{Basic setup and probability distributions}\label{subsec:distributions}

In this work, we address the mean gas density that photoionization modeling constrains. For this mean density $n_0$, the total column for a given ion $i$ of element $X$ can then be written as
\begin{equation}\label{eqn:column}
    N_{\rm X,i} = L n_0 A_{\rm X} Z \feff,
\end{equation}
where $L$ is the pathlength of the absorbing gas, $A_{\rm X}$ is the number density abundance in $X$ relative to hydrogen (for which we assume the \citealt{Asplund09} solar composition values), and $\feff$ is the effective ion fraction in the presence of a distribution in gas properties. For photoionized gas, the effective fraction is calculated by integrating over the temperature and density distributions, $\feff = \int{f_{\rm X,i}(n,T) g(n) p(T) n dn dT}$, where $\forg(n,T)$ is the ion fraction at a given gas density $n$ and temperature $T$, and $g(n)$ and $p(T)$ are the density and temperature probability distributions. In the scenarios we explore in this work the gas density and temperature are physically related. The ion fraction can then be written as a function of a single variable, and we take it to be the gas density, so that $f_{\rm X,i}(n,T(n))$. We now describe the density distributions we use, which we then plug into Equation~\eqref{eqn:eff_frac} to calculate the effective fractions.

\begin{figure}
\centering{\includegraphics[width=8cm]{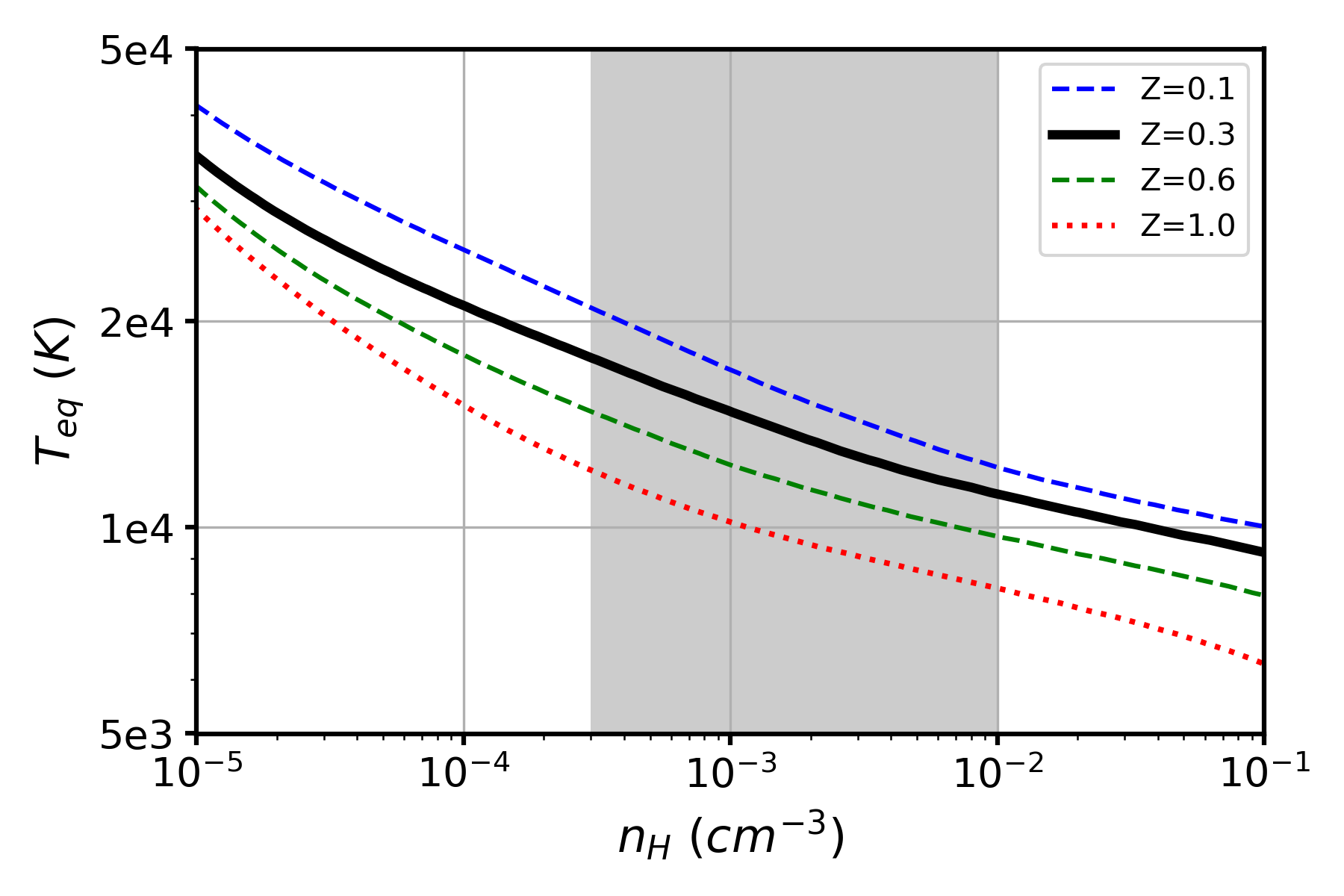}}
\caption{The equilibrium temperature at which photoheating balances cooling, $\teq$, as a function of density, and metallicity, calculated with the $z=0.2$ \citetalias{KS19} metagalactic radiation field. The grey shaded area indicates the range of densities inferred for the cool CGM by \citetalias{Werk14} and \citetalias{Prochaska17} (see also \S\ref{sec:fit_testing}). The temperature dependence on density, metallicity, and redshift can be approximated by power-law functions (see Equation~\ref{eqn:teq_fit}), and in \S\ref{subsec:redshift} we discuss how our analysis results vary with redshift.}
\label{fig:temp_eq}
\end{figure}

For the {\bf turbulent scenarios}, we assume a lognormal density distribution, given by 
\begin{equation}\label{eqn:lognormal_prob}
    g_{\rm turb}(n) = \frac{1}{n\sqrt{2\pi\sigma^2}} ~~ \exp{\left\{\frac{-\left[\ln(n/n_0)+\frac{\sigma^2}{2} \right]^2 }{2\sigma^2}\right \}} ~~~,
\end{equation}
as motivated in \S\ref{subsec:scen_turbulence}.
While the density distribution is the same for the isothermal and adiabatic cases, each scenario has a different temperature-density relation. For the {\bf isothermal case}, we assume the gas at each density is at the equilibrium temperature for that density, $\teq$. We plot $\teq$ in Figure~\ref{fig:temp_eq} as a function of gas density for metallicities between $0.1$ and $1.0$ solar. For the fiducial metallicity, $Z=0.3$ (shown by the solid black curve), the temperature can be approximated as $\teq \propto n_{\rm H}^{-0.145}$, accurate to within $15\%$ for $2 \times 10^{-5} < \nh/\cmv < 10^{-1}$.
While this is not exactly isothermal, $\teq$ is a weak function of the gas density, and we refer to this scenario as isothermal for simplicity. For the {\bf adiabatic case}, the temperature at the central density $n_0$ is $\teq$, and varies as $T \propto n^{2/3}$.

While in this work we perform our calculations at a constant metallicity and redshift, we note that more generally, the gas equilibrium temperature dependence on metallicity and redshift can be approximated as
\begin{equation}\label{eqn:teq_fit}
\begin{split}
    \teq \approx  1.35 \times 10^{4} \times &~ \left(\frac{n_{\rm H}}{10^{-3}~\cmv}\right)^{-0.145} \times \\
    &~ \left(\frac{Z}{0.3}\right)^{-0.188} (1+z)^{0.617} ~ {\rm K}
\end{split}
\end{equation}
for $0.1 \leq Z \leq 1.0$ and $0 \leq z \leq 0.5$. We discuss how our inferences for the CGM density scale with absorber redshift in \S\ref{subsec:redshift}. Figure~\ref{fig:temp_eq} shows the equilibrium temperature at $z=0.2$, calculated using \textsf{Cloudy} (without approximations) and plotted as a function of density for several metallicities.

For {\bf mixing layers}, we assume isobaric gas and that the probability to have gas at a temperature $T$ is independent of temperature, i.e. $p(T)$ is a constant, as found in numerical simulations ({\citealp{2019MNRAS.487..737J, Tan21a}; \S\ref{subsec:int_mixing}). As we show in Appendix~\ref{app_dist}, converting this temperature distribution to a probability as a function of density yields
\begin{equation}\label{eqn:mix_prob}
      g_{\rm mixing}(n) = \frac{\fmm}{\ln{\left(\thot/\teq\right)}} \frac{1}{n^2} ~~~ \text{for}~~~ n < n_0 ~~~,
\end{equation}
where $\fmm$ is the mass ratio of mixing to cool gas and $\thot$ is the temperature of the hot phase. This distribution extends down to densities of $n_0\; \teq/\thot$, and our calculations are weakly sensitive to the choice of the maximum temperature (or minimum density).

For {\bf cooling gas}, we assume that the cooling is isobaric and that the probability to have gas at temperature $T$ is proportional to its cooling time, $p(T) \propto T/ n \Lambda$ (\citealp{Heckman02, Qu18a}; \S\ref{subsec:int_cooling}). As we show in Appendix~\ref{app_dist}, the probability as a function of density can be written as
\begin{equation}\label{eqn:cool_prob}
      g_{\rm cooling}(n) =  \frac{\fmm}{I_n(\teq,T_{\rm warm})} \frac{1}{\Lambda n^4}  ~~~ \text{for}~~~ n < n_0 ~~~,
\end{equation}
where $\fmm$ is the mass ratio of cooling to cool gas, $I_n(\teq,T_{\rm warm})$ is the integral of the PDF over the relevant density (or temperature) range (see below), and $\Lambda$ is the gas cooling efficiency.

Unlike the PDF in the mixing layers model (Eq.~\ref{eqn:mix_prob}), the PDF in Eq.~\ref{eqn:cool_prob} has a strong dependence on the gas density (and temperature), with most of the gas mass residing at low densities (or equivalently high temperatures), making the normalization sensitive to the cutoff value. Since we aim to model gas that is actively cooling, we normalize the PDF by the integral between $\teq$ and $T_{\rm warm} = 2 \times 10^5$~K. As discussed briefly in \S~\ref{subsec:int_cooling}, this normalization connects to our estimates for the value of $f_M$, based on mass accretion rates and \OVI\ columns. The integral over the full temperature range, up the hottest temperature that is assumed to exist in the cooling gas PDF, $T_{\rm hot},$ gives the total warm/hot to cool gas mass ratio, $M_{\rm hot}/\mcool$\footnote{We note that $\teq$ is a function of the cool gas density (see Figure~\ref{fig:temp_eq} and Eq.~\ref{eqn:teq_fit}), and as a result, the value of the integral over the full PDF also varies weakly as a function of $n_0$.}. The effective fractions of high ions in this model show a mild dependence $T_{\rm hot}$. Our choice for this work, $T_{\rm hot} = 10^6$~K, is motivated by the virial temperature of a MW-mass halo, and we note that higher values produce $M_{\rm hot}/\mcool$ values that are inconsistent with ratios estimated for the multiphase CGM of MW-like galaxies (see Appendix~\ref{app_dist}).

To summarize, the density-temperature, $T(n)$, relations for the scenarios we consider are
\begin{equation}\label{eqn:tn_rel}
\begin{split}
T(n) = \begin{cases}
& \teq(n) \approx \teq(n_0) \times \left(n/n_0\right)^{-0.145} ~~ \text{isothermal turb.};\\
& \teq(n_0) \times \left(n/n_0\right)^{2/3} ~~ \text{adiabatic turbulence}; \\
& \teq(n_0) \times \left(n/n_0\right)^{-1} ~~ \text{isobaric cooling/mixing}.
\end{cases}
\end{split}
\end{equation}
For isothermal turbulence, the density scaling written here is approximate; our calculations use the full numerical results for $\teq(n)$ from \textsf{Cloudy}.

Next, we use these probability distributions to calculate the effective ion fractions.

\subsection{Effective ion fractions}\label{subsec:effective_fractions}

We use the \textsf{Cloudy 17.00} code \citep{Ferland17} to obtain the ion fractions, $\forg$, as functions of gas temperature and density for photoionized gas in the presence of MGRF. We adopt the \citet[hereafter KS19]{KS19} MGRF and perform our calculations at $z=0.2$, the median galaxy redshift in the COS-Halos data set. Since the shape of the MGRF spectrum remains approximately the same at $z<2$, our results can be easily scaled to other redshifts (see \S\ref{subsec:redshift}).  

Given the ion fractions, density distribution, $g(n)$, and temperature-density relation, $T(n)$, for each physical scenario, we can then calculate the (mass-weighted) effective ion fraction
\begin{equation}\label{eqn:eff_frac}
    \feff = \frac{\int{\forg(n,T) \,n \,g(n) \,dn}}{\int{n \,g(n) \,dn}} ~~~.
\end{equation}
The resulting effective fraction depends on $n_0$ and the properties of the distribution. For the mean densities, we examine the range $10^{-5}<n_0/{\rm cm}^{-3}<10^{-1}$, easily encompassing the cool gas densities in the COS-Halos sample \citep[hereafter W14]{Werk14}. We consider only optically thin gas and address how self-shielding by \HeII\ affects our results in \S\ref{subsec:HeII_SS} and Appendix~\ref{app_HeII}. As discussed in \S\ref{sec:scenarios}, we parameterized the distributions with a single parameter in each case. For the turbulent models, this is the width of the lognormal distribution, $\sigma$, and we address the range between $\sigma = 0$ (no fluctuations) and $1.2$, a range motivated by the estimates in \S\ref{subsec:scen_turbulence}. For the cooling and mixing scenarios, the parameter is $\fmm$, the mass fraction in cooling or mixing gas relative to the mass in the cool phase, and we consider the range between $\fmm = 0$ (no mixing/cooling) and $0.6$, a range motivated in \S\ref{subsec:scen_int}. For each of the scenarios we address, this calculation results in the effective ions fractions as functions of the central or cool gas density and model parameter, $\feff(n_0,\sigma)$ or $\feff(n_0,\fmm)$.

\begin{figure*}\centering{
\includegraphics[width=0.99\textwidth]{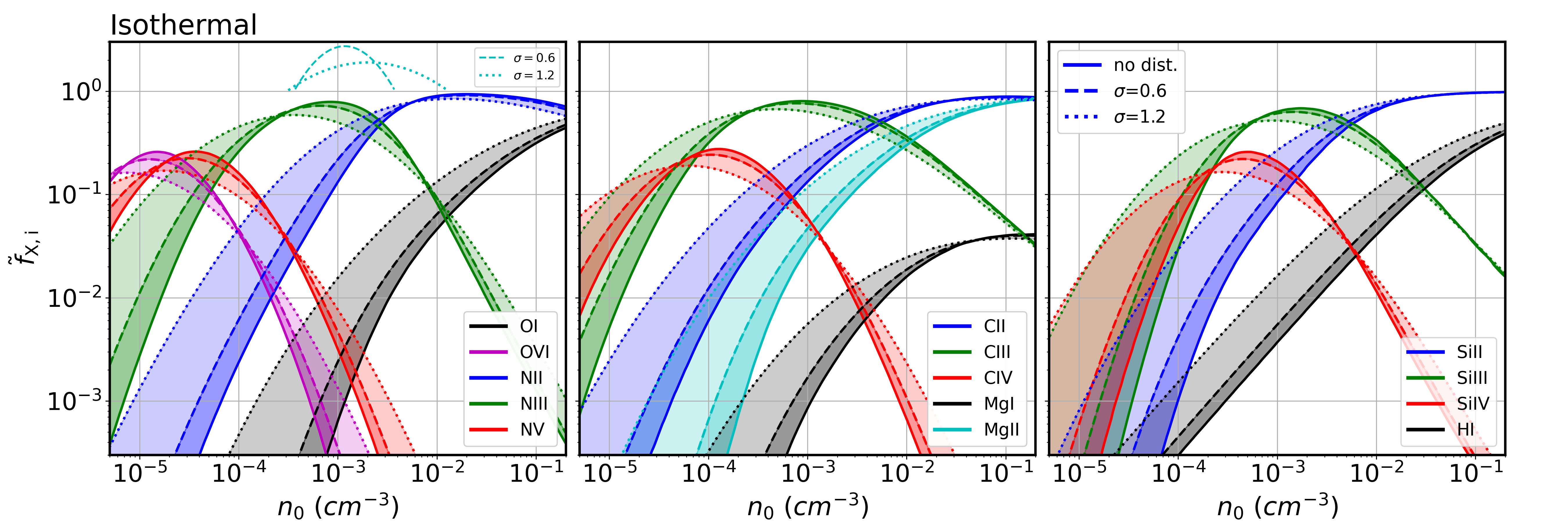}
\includegraphics[width=0.99\textwidth]{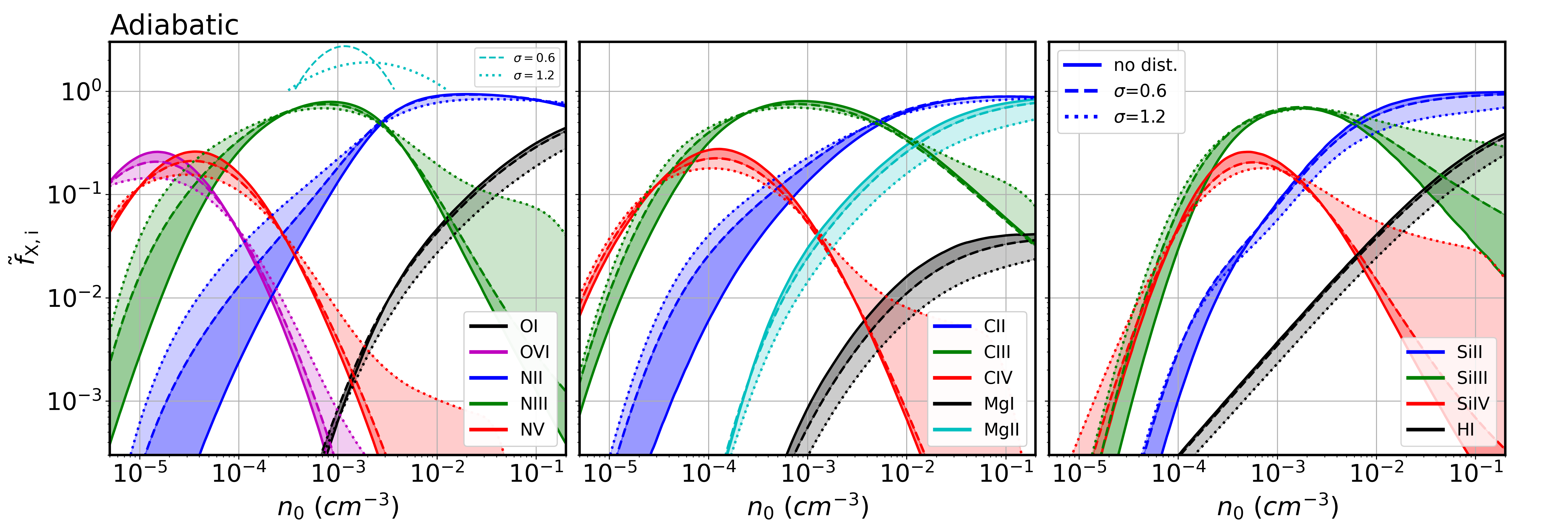}}
\caption{Effective ion fractions for the lognormal density and temperature fluctuations, for the {\bf isothermal and adiabatic turbulence} models (top and bottom, respectively). The solid curves show the fractions without fluctuations, and dashed and dotted curves are for $\sigma=0.6$ and $\sigma=1.2$, respectively, motivated by the velocity widths of CGM clouds (see \S~\ref{subsec:scen_turbulence} and \S~\ref{subsec:effective_fractions}). To emphasize the change in ion fraction, the dark and light shading show the difference between the baseline model and $\sigma=0.6$, and between $\sigma=1.2$ and $\sigma=0.6$, respectively. Here we use the ionizing background model of \citetalias{KS19} and $n_0$ is the mean hydrogen density if the system is at $z=0.2$, but can be easily rescaled to different redshifts through the ionization parameter (see \S\ref{subsec:redshift}). The dashed cyan curves in the left panels show the mass-weighted lognormal PDFs ($n^2g(n)$) on a logarithimic y scale, for illustration, with a mean value of $n_0 = 10^{-3}~\cmv$.}
\label{fig:eff_frac1}
\end{figure*}

We plot the results for the effective ion fractions in Figure~\ref{fig:eff_frac1} for the turbulent scenarios and in Figure~\ref{fig:eff_frac2} for cooling and mixing gas. We show \HI~and a set of metal ions that is motivated by observations -- \CII-\CIV, \NIII-\NV, \OI, \OVI, \MgI, \MgII, and \SiII-\SiIV. We also provide the model outputs as tables attached to this manuscript for use in future works.

Figure~\ref{fig:eff_frac1} shows the ion fractions for the isothermal and adiabatic turbulence scenarios (top and bottom rows, respectively) as functions of the mean density $n_0$. The solid lines show the effective fractions for no fluctuations, while the dashed and dotted curves are for $\sigma=0.6$ and $\sigma=1.2$, respectively. The cyan curves in the top left panel shows the distributions for a model with $n_0 =3\times10^{-3}$. In general, for both the isothermal and adiabatic cases, the peak value of each ion fraction, at its characteristic gas density, is only weakly affected by the fluctuations. The relative effect of the fluctuations is larger at densities far from this peak, where the ion fraction without fluctuations is low and any (absolute) small increase in ion fraction makes a larger difference.

As an example, we examine the \CII\ fraction, plotted by the blue curves in the middle panels. At $\nh \sim 3 \times 10^{-2}~\cmv$, at the limit of the density range relevant for the CGM, the fraction without fluctuations (solid curve) is close to unity over a range of densities and the density distribution does not significantly affect the effective fraction, $\tilde{f}_{\rm CII}$ (dashed and dotted). At lower densities and for a wide distribution ($\sigma = 1.2$), the \CII\ fraction is significantly boosted -- at $\nh \approx 10^{-3}$ and $10^{-4}~\cmv$ it is higher than the no fluctuation case by factors of $\sim 3$ and $\sim 10$, respectively. This is a result of the $f_{\rm CII}(n, \teq)$ increasing strongly with density, so that the high density tail of the distribution brings in substantially more \CII. 

It is interesting to note that the increase in \CII\ at $\nh \approx 10^{-4}~\cmv$, for example, is similar for $\sigma=0.6$ and $\sigma=1.2$ for the adiabatic case (both with a factor of $\approx 10$ enhancement relative to $\sigma=0$), but much more modest for the isothermal $\sigma=0.6$ case, with a factor of $\sim 2$ boost in the ion fraction. This is a result of the difference in gas temperature variation -- for the isothermal case, only the density leads to a change in the ion fraction, and wide distributions are needed for significant variation in $\feff$. For the adiabatic case, the gas temperature varies more strongly with density (see Eq. \ref{eqn:tn_rel}), leading to an ion fraction curve that is more strongly peaked at a specific density. Then, the peak being ``picked up'' by the PDF (even for lower $\sigma$) leads to a sharp increase from $\sigma=0$.

The behavior of the effective ion fractions is similar for ionization states similar to \CII, such as \NII\ (left panels), \MgII\ (middle), and \SiII\ (right). It is also qualitatively similar for even lower ionization states (such as \HI\, \MgI, and \OI), with peak fractions at densities higher than the values we consider here.

For higher ions, with peak fractions at densities around $10^{-3}~\cmv$, the peak ion fraction also does not vary significantly, with small and negligible decreases for the widest distributions in the equilibrium and adiabatic cases, respectively. At densities above and below the peak, the ion fraction increases in the presence of a lognormal distribution. As noted, the fractions of these ions are naturally low at higher densities, and even a small (absolute) increase resulting from the distribution ``picking up'' lower densities can be significant.

For the isothermal scenario, the ion fractions are boosted more significantly at lower densities, since this is where the ion fraction decreases more rapidly with decreasing density (similar to the behaviour of low ions described earlier). For the adiabatic case, there is a larger increase in $\feff$ at higher densities. This is a result of the temperature variation in the wing of the lognormal (as $T \propto n^{2/3}$) leading to a stronger decrease in ion fraction at higher densities (compared to isothermal or baseline).
This is especially prominent for \SiIV~and \CIV, ions that without fluctuations have low fractions at $\teq$ and $\nh>10^{-2}~\cmv$ (solid curves). However, their fractions peak at $T\approx 10^5$~K due to collisional ionization, and for $\sigma = 1.2$ these temperatures make a significant contribution to the effective fraction at high (mean) densities.

\begin{figure*}\centering{
\includegraphics[width=0.99\textwidth]{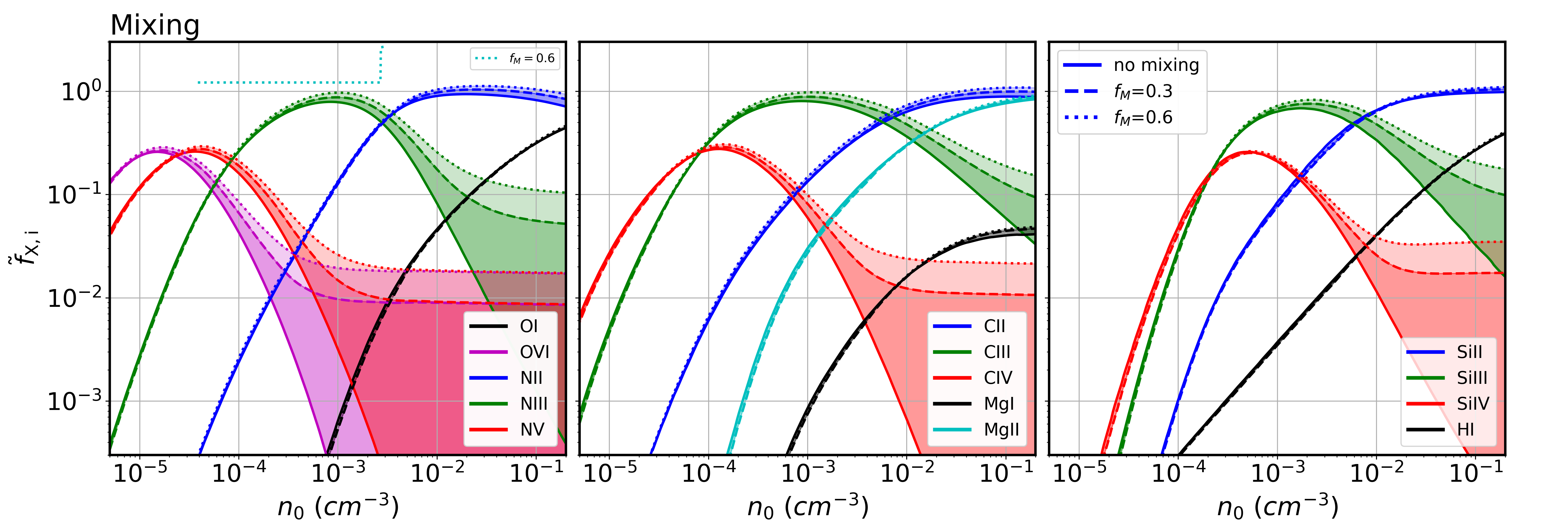}
\includegraphics[width=0.99\textwidth]{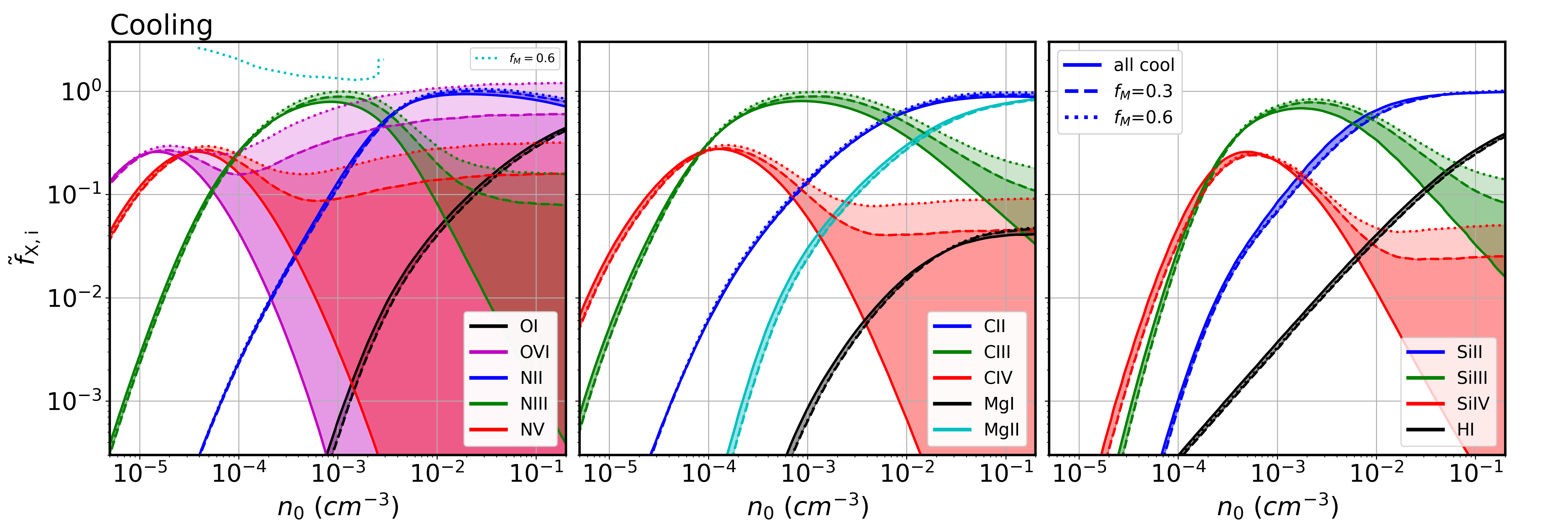}}
\caption{Effective mass-weighted ion fractions for the {\bf mixing layers} (top) and {\bf cooling gas} (bottom) scenarios, with isobaric distributions. The solid curves show the fractions without mixing or cooling gas, and dashed and dotted curves are respectively for mass fractions of $\fmm=0.3$ and $0.6$ in intermediate temperature gas (motivated by cooling rates, see \S\ref{subsec:scen_int} and \S\ref{subsec:effective_fractions}). These calculations are for $z=0.2$, but can be easily rescaled to different redshifts through the ionization parameter (see \S\ref{subsec:redshift}). The dotted cyan curves in the left panels show the PDF shape, plotting $n^2g(n)$ on a logarithimic y scale, for illustration, for $\fmm=0.6$ and $n_0 = 3 \times 10^{-3}~\cmv$. The shape for $\fmm=0.3$ is identical, only varying by a factor for gas at $n<n_0$.}
\label{fig:eff_frac2}
\end{figure*}

Figure~\ref{fig:eff_frac2} shows the ion fractions for mixing and cooling gas scenarios as a function of the density of cool gas $n_0$ (top and bottom rows, respectively). The solid lines show the original fractions, without mixing or cooling gas, while the dashed and dotted curves are for $\fmm=0.3$ and $\fmm=0.6$, respectively. These values are chosen to saturate the amount of cooling gas that is possible per our estimates in \S~\ref{subsec:int_mixing} and \ref{subsec:int_cooling}. The cyan curves at the top of the left panels shows the density distributions for a model $n_0 =3\times10^{-3}~\cmv$. The mixing/cooling gas is at higher temperatures, and in these isobaric models the distributions extend to densities lower than that of the cool gas. For both models, the effect of the gas density distributions on the mixing and cooling gas ion fractions is negligible at or close to their peak densities, similar to the turbulent model. This is also true for most ions at densities below the peak. This can be explained by the fact that for isobaric gas, lower densities have higher temperatures, and ions are removed both by collisions and radiation, leading to no contribution to that ion from cooling and mixing layers. On the other hand, for densities above the peak ion fraction, cooling and mixing gas can lead to a significant increase in the ion fraction, compared to the no fluctuations case. This is especially true for the higher ions, such as \SiIV~(red curve in the right panels), \CIV~(middle), and \NV~and \OVI~(red and magenta curves in the left panels, respectively). The reason is that even if the gas densities in mixing/cooling gas are too high for radiation to allow these ions to form through photoionization, the higher temperatures of the isobaric distribution contribute to the effective ion fractions through collisional ionization.
For example, for $\fmm=0.6$, in the {\bf mixing layers} model the ion fractions of \NV\ and \OVI\ can be $\approx (0.3-3) \times 10^{-2}$. However, even with these high (relative) boosts, the effective ion fractions are still a factor of $30-100$ below their peak values at lower densities, suggesting that the total absorption columns may be insufficient for detection. In the {\bf cooling gas model}, these ions have high effective fractions, of $0.1-1$, similar to or above the fractions at the photoionization peak density. These high fractions are a result of a fraction of $\approx f_M$ of the gas mass (relative to the cool gas) residing at intermediate temperatures of $T\sim 10^5$~K in these models -- and even more at higher temperatures-- where collisions can set the ionization and the ion fractions of \CIV, \NV, and \OVI\ are high.

While for the low ions the predictions of mixing layers and cooling gas models are similar, they differ in their \CIV, \NV\ and \OVI\ ion fractions, and we discuss these differences in the next section when comparing our models to observations.

\section{Testing models against observations}
\label{sec:fit_testing}

We have so far investigated the impact of a lognormal distribution and intermediate temperature gas models on the ion fractions. We now test whether these effects may be observable or even favored by the data by comparing them to the COS-Halos measurements. We focus on this data set as it is the most uniform and complete sample probing the inner CGM of $L^*$ halos at low redshift, and all the data and detailed fits are available online. We first describe our sample selection and fitting method and then present the results of our analysis.

\subsection{Sample selection and fitting method}
\label{subsec:fit_method}

\begin{table*}
\centering
\caption{Summary of the fits to sightlines for the different models and scenarios we consider and the model parameters ($\sigma$ or $\fmm$). Columns show the COS-Halos sightline label, the number of metal lines with columns detected or limits reported, and the density (and the $\chi^2$ value) of the best-fit model. We present here a partial list of the results for brevity, and provide the full table, including all model parameter values, density uncertainties, and the inferred metallicities, in the data file attached to the manuscript. See Figures~\ref{fig:fits_lognormal}-\ref{fig:fits_mix} for the fits (measured and model column densities), and Figures~\ref{fig:nh_compare}-\ref{fig:nhi_crit} for a summary of the densities and metallicities (including uncertainties).}

\label{tab:fit_results}
\begin{tabular}{| l | c  c | c | c c | c | c | c | c c |}
\hline
sightline & $N_{\rm det}$ & $N_{\rm tot}$  & \multicolumn{8}{|c|}{$\nh~(10^{-3}~\cmv)~~~(\chi^2)$} \\
\hline
& &  & \texttt{baseline}
& \multicolumn{2}{|c|}{\texttt{isothermal}} 
& \texttt{adiabatic}
& \texttt{mixing}
& \texttt{cooling}
& \multicolumn{2}{|c|}{\texttt{self-shielded}} \\
& & & & $\sigma =0.6$ & $\sigma =1.2$ 
& $\sigma =1.2$ & $\fmm =0.6$ & $\fmm =0.6$ & $\sigma =0$ & $\sigma =1.2$ \\
\hline

\multirow{2}{*}{J0910+1014\_242\_34} & \multirow{2}{*}{4} & \multirow{2}{*}{7}
& $3.6$ & $2.5$ & $0.98$ & $8.7$ & $4.5$ & $4.4$ & $3.2$ & $0.83$ \\
& & & $(8.3)$ & $(9.6)$ & $(11.5)$ & $(11.6)$ & $(8.5)$ & $(7.9)$ & ($15.8$) & ($20.3$) \\ 

\multirow{2}{*}{J1016+4706\_274\_6} & \multirow{2}{*}{4} & \multirow{2}{*}{9}
& $1.9$ & $1.8$ & $0.60$ & $2.9$ & $2.5$ & $2.4$ & $1.2$ & $0.37$ \\
& & & $(16.3)$ & $(14.1)$ & $(8.0)$ & $(49.1)$ & $(19.4)$ & $(18.4)$ & ($0.3$) & ($0.3$) \\

\multirow{2}{*}{J1233+4758\_94\_38} & \multirow{2}{*}{4} & \multirow{2}{*}{8}
& $2.3$ & $1.8$ & $0.74$ & $3.0$ & $2.7$ & $2.5$ & $1.0$ & $0.44$ \\
& & & $(6.8)$ & $(6.4)$ & $(5.7)$ & $(21.1)$ & $(6.8)$ & $(5.6)$ & ($16.3$) & ($6.4$) \\ \hline

\multirow{2}{*}{J1241+5721\_208\_27} & \multirow{2}{*}{3} & \multirow{2}{*}{8}
& $0.94$ & $0.83$ & $0.26$ & $1.5$ & $0.94$ & $1.3$ & $0.33$ & $0.15$ \\
& & & $(84.7)$ & $(76.0)$ & $(66.7)$ & $(119.7)$ & $(98.4)$ & $(97.0)$ & ($106.3$) & ($72.5$) \\

\multirow{2}{*}{J1322+4645\_349\_11} & \multirow{2}{*}{4} & \multirow{2}{*}{9}
& $1.7$ & $1.2$ & $0.44$ & $2.0$ & $2.0$ & $1.9$ & $0.94$ & $0.26$ \\
& & & $(2.7)$ & $(2.0)$ & $(1.3)$ & $(18.7)$ & $(3.3)$ & $(3.5)$ & ($7.4$) & ($8.0$) \\

\multirow{2}{*}{J1330+2813\_289\_28} & \multirow{2}{*}{5} & \multirow{2}{*}{10}
& $3.7$ & $4.0$ & $2.0$ & $6.3$ & $5.3$ & $4.7$ & $2.6$ & $1.2$ \\
& & & $(59.1)$ & $(53.3)$ & $(39.4)$ & $(82.2)$ & $(58.5)$ & $(65.5)$ & ($40.1$) & ($35.2$) \\ \hline

\multirow{2}{*}{J1550+4001\_197\_23} & \multirow{2}{*}{4} & \multirow{2}{*}{9}
& $1.2$ & $0.74$ & $0.19$ & $1.6$ & $1.4$ & $1.5$ & $0.60$ & $0.15$ \\
& & & $(12.9)$ & $(11.9)$ & $(8.7)$ & $(29.4)$ & $(12.2)$ & $(13.3)$ & ($17.9$) & ($16.3$) \\

\multirow{2}{*}{J1555+3628\_88\_11} & \multirow{2}{*}{6} & \multirow{2}{*}{9}
& $1.4$ & $0.94$ & $0.32$ & $1.7$ & $1.5$ & $1.6$ & $0.60$ & $0.23$ \\
& & & $(11.8)$ & $(10.0)$ & $(9.4)$ & $(25.1)$ & $(13.8)$ & $(12.1)$ & ($46.1$) & ($24.8$) \\

\multirow{2}{*}{J2345-0059\_356\_12 } & \multirow{2}{*}{3} & \multirow{2}{*}{6}
& $0.56$ & $0.32$ & $0.065$ & $0.37$ & $0.60$ & $0.060$ & $0.21$ & $0.058$ \\
& & & $(3.2)$ & $(3.7)$ & $(4.9)$ & $(5.0)$ & $(3.2)$ & $(3.2)$ & ($23.6$) & ($19.5$) \\

\hline
\end{tabular}
\end{table*}

We compare our calculations to the measurements collected as part of the COS-Halos survey, reported by \citet[hereafter W13]{Werk13} (see also \citealp{Tumlinson11}, \citealp{Werk12}, \citetalias{Werk14}). The survey used UV absorption spectroscopy of background QSOs to study the CGM of galaxies residing in halos with $M_{\rm vir} \sim 10^{12}M_\odot$ (given by stellar-mass abundance matching methods), similar to the Milky Way, at redshifts $0.1<z<0.4$ (median of $z\approx 0.2$). Our choice of the COS-Halos data set is motivated by the size of the sample and its coverage of a wide range of metal atomic lines, with several different ions detected per system. Another advantage of this data set is that the majority of the COS-Halos absorption systems have \HI\ columns for which hydrogen self-shielding is small or negligible (see \citealp[hereafter P17]{Prochaska17}), allowing us to avoid modeling the more complex radiative transfer physics (although see our discussion on \HeII\ self-shielding in \S\ref{subsec:HeII_SS}).

We test our models, that allow for distributions in density and temperature, against singe-density, equilibrium-temperature models. For this purpose, we apply two selection criteria to the COS-Halos sample. First, we restrict to systems that have at least three metal line column measurements (i.e. that are not upper or lower limits), since we are fitting for two parameters -- density and total metal column -- while a third parameter describing our model (e.g. $\sigma$ in our lognormal models) is not fit for explicitly\footnote{We have applied our fitting procedure to the larger set of absorption line systems that have fewer than three detections -- in all but one such system, we find that systems with fewer than three detected lines yield extremely small $\chi^2$ values, indicating over-fitting.}. Second, we only consider systems that have $N_{\rm HI} < 10^{17.5}~\cmc$, below the \HI\ column where the neutral hydrogen starts to self-shield, given by optical depth of unity at $1$~Ry. This allows us to compute ionization fractions in the optically thin limit (with the complication that \HeII\ may self-shield, which we address in \S\ref{subsec:HeII_SS} and in Appendix~\ref{app_HeII}). Applying these two cuts leaves nine COS-Halos systems. In addition to a minimum of three detections, these lines of sight have upper or lower limits on the column densities of additional ions, which we also use in our analysis.

We generally fit the total columns reported in \citetalias{Werk13}. Our modeling implicitly assumes the absorption of different ions is co-spatial and hence should appear at similar velocities\footnote{We note that while kinematic alignment of absorbers does not necessarily indicate spatial alignment, we require the former as a necessary, even if possibly insufficient condition for co-spatiality.}. We have made several minor corrections to bring our systems to accord with this principle. In particular, in system J0910+1014\_242\_34 there is one \CII\ absorption component with a velocity well offset from the absorption in other ions. We removed this velocity component from the total inferred column, leaving $N_{\rm CII} = 10^{13.89}~\cmc$ for our fit. For similar reasons, we removed some of the \SiII\ and \SiIII\ components from the total columns in J1322+4645\_349\_11 leaving $N_{\rm SiII}= 10^{13.37}~\cmc$ and $N_{\rm SiIII}= 10^{13.73} \pm 0.2~\cmc$. We note the error in the latter is the result of removing a saturated line, converting a bound to a constraint. Finally, for J1550+4001\_197\_23, we altered the \MgII\ and \CII\ columns such that $N_{\rm MgII}=10^{12.63}~\cmc$ and $N_{\rm CII}= 10^{13.98}~\cmc$.

Using our models for the effective ion fractions, we can calculate $N_{\rm X,i}$, the total model column density of ion $i$ of element $X$, given by Equation~\eqref{eqn:column}.
Then, to fit each model, we vary the total system metal ion column and density in order to minimize $\chi^2$, defined as
\begin{equation}\label{chi2}
\chi^2 = \sum_{X,i \in \rm ions}  \frac{\left(\log_{10}(N_{\rm X,i}) - \log_{10}(\widehat{N}_{\rm X,i})\right)^2}{\sigma_{N_{\rm X,i}}^2} ~{\cal B}_{\rm X,i},
\end{equation}
where $N$ is the model column density that depends on the gas density and the pathlength in metals given by $Z L$, $\widehat{N}$ is the measured column density, $\sigma_{N}$ is the column measurement uncertainty, and $\cal B$ is a function that is related to the limits in the data and we describe below. The sum goes over all observed ions that were not marked as blends with other lines in \citetalias{Werk13}, after we exclude high ions and low ions from our fits, namely any metal in the atomic phase, \NV\ and \OVI. While \OVI\ likely owes to relatively hot gas, \NV\ may arise in the photoionized phase. There is only one detection of \NV\ in our sample, and otherwise bounds. We address later how our results are affected if we include it. We also do not include iron ions in our modeling.

For column density measurements (not limits), the error $\sigma_{N}$ is either the error reported in \citetalias{Werk13} for $\log_{10}(\widehat{N}_{\rm X,i})$ or $\sigma_{N,{\rm min}}=0.1$, the larger of the two. Since the errors reported by \citetalias{Werk13} are in the range $0.05-0.1$ for most of the detected lines, our approach sets $\sigma_{N} = 0.1$ for the vast majority of column measurements. Thus, our goodness-of-fit results can be rescaled to a different average error $\sigma$ approximately as $\chi^2 \rightarrow \chi^2 \times (0.1/\sigma)^2$. The function ${\cal B}_{\rm X,i}$ is introduced to deal with limits, and for measurements, ${\cal B}_{\rm X,i} = 1$. For upper bounds, ${\cal B}_{\rm X,i} = \Theta(\widehat{N}_{\rm X,i}' - N_{\rm X,i})$, and similarly ${\cal B}_{\rm X,i} = \Theta(N_{\rm X,i} -\widehat{N}_{\rm X,i}')$ for lower bounds, where $\Theta$ is the Heaviside function. For bounds, $\widehat{N}_{\rm X,i}'$ is the value of the bound reported in \citetalias{Werk13} and in Equation~\eqref{chi2} we set $\widehat{N}_{\rm X,i} = \widehat{N}_{\rm X,i}'\pm 2\sigma_{N}$ where again $\sigma_{N}=0.1$ and where the minus is for an upper bound and plus for a lower bound.  This results in nothing being added to $\chi^2$ if the model is within the bound, but if it is at the bound the penalty in $\chi^2$ is the same as being two standard deviations off, and this penalty of course increases as the bound is further violated.

We perform the fitting by brute force: scanning over the full range of densities, with $n_0 \in 10^{-5} - 10^{-1}~\cmv$ (shown in the previous plots), and the metallicity-pathlength range, $Z L \in \{10^{16}-10^{22}\}/n_0$~kpc, which we find the best-fit is always well within. Owing to our treatment of bounds, $\chi^2$ is discontinuous and so gradient descent-like minimization algorithms are not possible. In order to calculate the 1$\sigma$ standard deviation on $n_0$, once we find the best-fit model, we then vary first $n_0$ from the best-fit value until $\min_{ZL}[\chi^2(n_0)]$ changes by $2.3$, where $\min_{ZL}$ finds the minimum over all $ZL$. The value $2.3$ is set by the contour in a 2-parameter Gaussian that contains $68\%$ of the area. In addition to scanning over $\nh$ and $ZL$, we also scan over the parameter that controls the density and temperature distribution, such as the width of the distribution for lognormal ($\sigma$) or the amount of gas in intermediate temperature phase for mixing and cooling ($\fmm$) as detailed in \S\ref{sec:scenarios}. For this scan over model parameters, we only consider a handful of values and report the best fit in density for each model parameter and scenario (see \S\ref{subsec:fit_results} and Table~\ref{tab:fit_results}).

Finally, this fitting procedure does not use constraints on the \HI\ column density.  The \HI\ column density allows one to constrain metallicity $Z$, but does not aid in constraining $n_0$\footnote{There is a small caveat here, that the metallicity affects $\teq$ which in turn modestly affects $\forg$. However, as discussed in \S\ref{subsec:distributions} and shown in Figure~\ref{fig:temp_eq}, this dependence is weak, with a factor of $\approx 30\%$ difference in temperature for a $1$~dex change in metallicity, between $Z=0.1$ and $1$. In this work we ignore this additional dependence and assume a nominal value of $Z=0.3$ for the ion fraction calculations. In Appendix~\ref{app_met_var} we show explicitly how some selected ion fractions vary with the metallicity (see Figure~\ref{fig:metallicity}).}. Thus, once we incorporate $\widehat{N}_{\rm HI}$, we can infer the system's metallicity as 
\begin{equation}
\widehat{Z} =  \frac{\widehat{Z L}~ \widehat{n}_{0}}{\widehat{N}_{\rm HI}/f_{\rm HI,i}|_{\rm BF}},
\end{equation}
where $f_{\rm HI,i}|_{\rm BF}$ indicates the fraction evaluated at the best fit density, and the metallicity errors are calculated by propagating the errors on $\widehat{N}_{\rm HI}$, $\widehat{n}_{0}$ and $\widehat{Z L}$.

We now describe the results of our analysis.

\begin{figure*}
{\centering{\hspace*{-0.6cm}\includegraphics[width=1.1\textwidth]{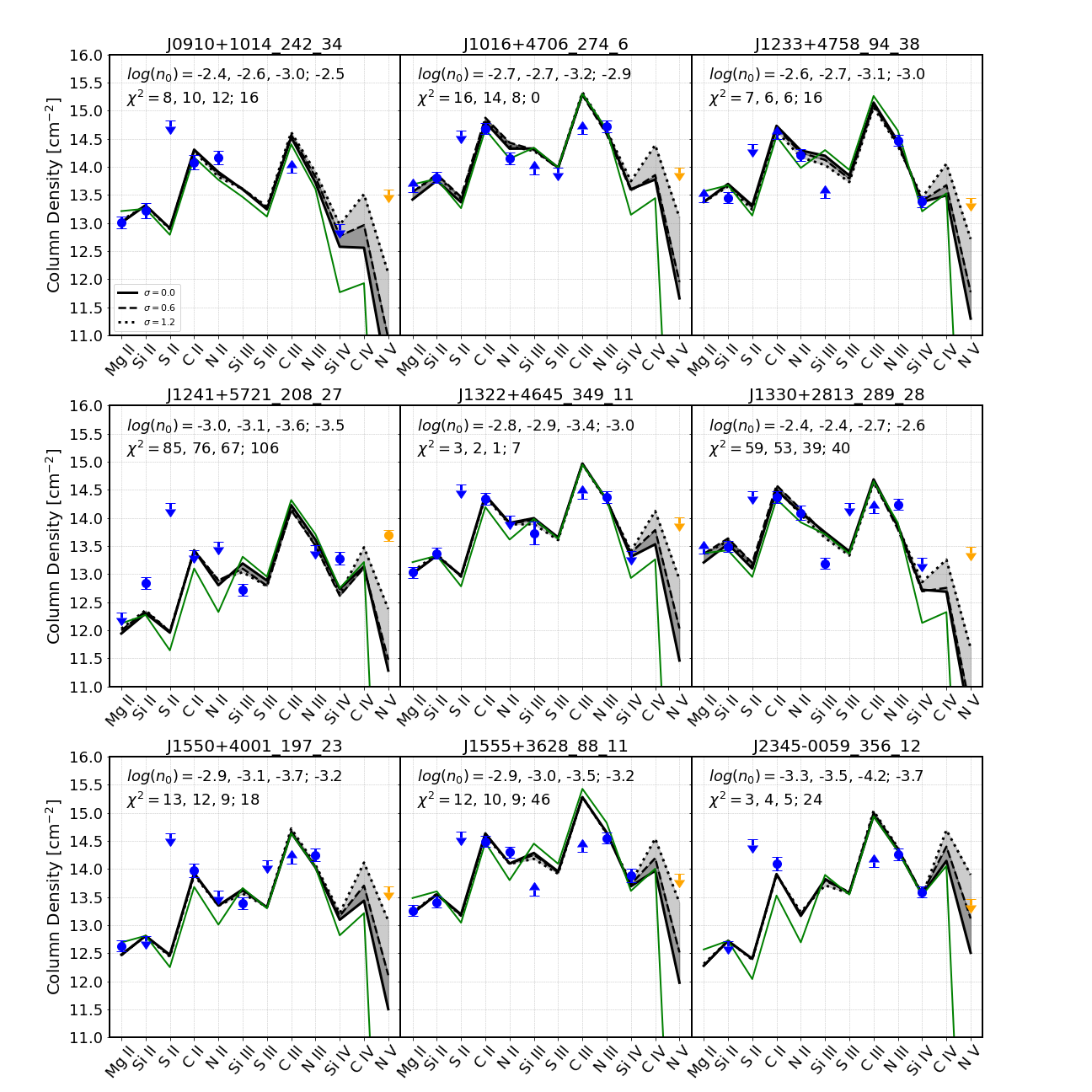}}}
\vspace{-0.3cm}
\caption{Comparison of the {\bf isothermal turbulence model} outputs (curves)  with the COS-Halos data (markers). The ions are ordered by increasing ionization energy. Each panel shows an object in our sub-sample, selected from \citet[see \S\ref{subsec:fit_method}]{Werk12}. Thick black curves are for optically thin gas and lognormal density distributions with $\sigma = 0$ (no fluctuations), $0.6$, and $1.2$ (solid, dashed, and dotted curves, respectively, similar to the styling in Figures~\ref{fig:eff_frac1}-\ref{fig:eff_frac2}). Thin solid green curves show a model with full \HeII\ self-shielding and $\sigma=0$, which we discuss in \S\ref{subsec:HeII_SS}. The text in the upper left corner indicates the density and $\chi^2$ values for the four models in the order described (see also Table~\ref{tab:fit_results}). The yellow markers for \NV\ indicate that it was not used in the fits. The only systems where \NV\ data constrain the models are 208\_27 and 356\_12, and we discuss these in \S\ref{subsec:fit_objects}.
\label{fig:fits_lognormal}}
\end{figure*}

\begin{figure*}
{\centering{\hspace*{-0.4cm}\includegraphics[width=1.1\textwidth]{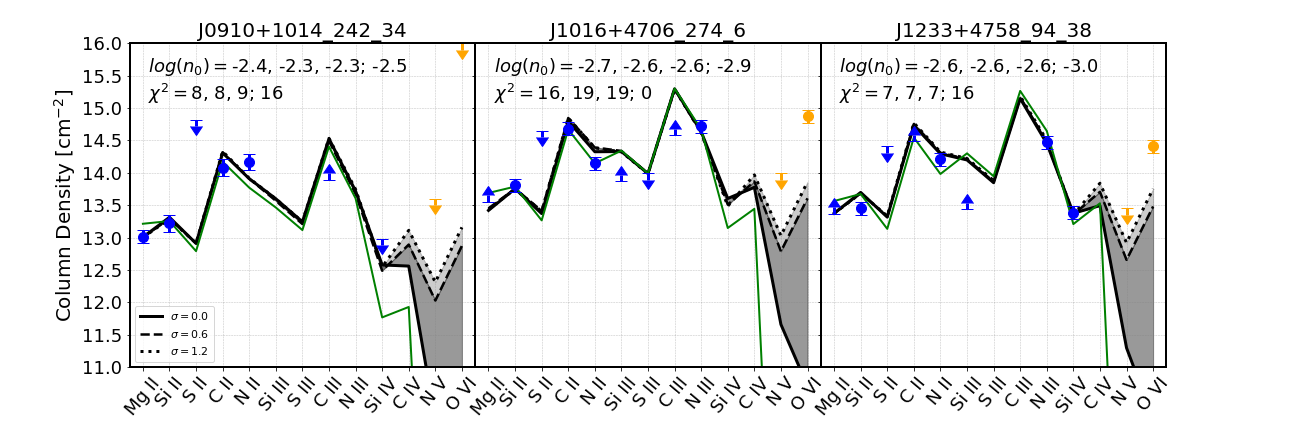}}}
\vspace{-0.3cm}
\caption{Same as Figure~\ref{fig:fits_lognormal} {\bf but for the mixing layers scenario}, with $\fmm=0$ (identical to the $\sigma=0$ baseline model), $\fmm=0.3$, and $\fmm=0.6$ (thick black solid, dashed, and dotted curves, respectively), for the first three objects in our sample. As discussed in \S\ref{subsec:effective_fractions}, the effective ions fractions and column densities of high ions are boosted significantly even for small $\fmm$, with a linear increase for higher values. Orange points are relevant higher ions that are not included in the fits. For reasonable values of $\fmm$ the model produces \OVI\ columns significantly lower than measured (which could be explained for example by this column owing to virialized gas). The thin solid green curve is the self-shielded model with $\fmm=0$.}
\label{fig:fits_mix}
\end{figure*}

\begin{figure*}
{\centering{\hspace*{-0.4cm}\includegraphics[width=1.1\textwidth]{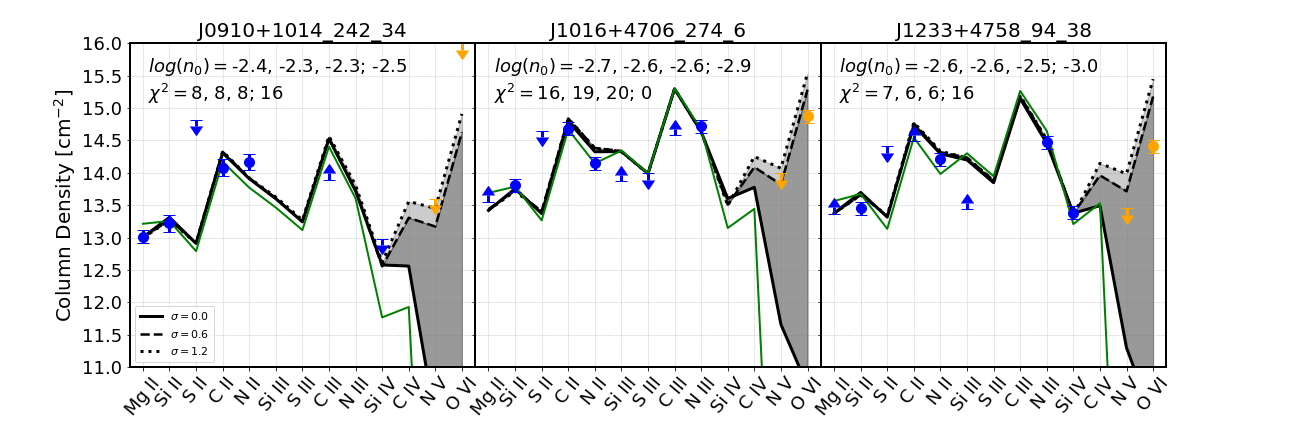}}}
\vspace{-0.3cm}
\caption{Same as Figure~\ref{fig:fits_mix} {\bf but for the cooling gas scenario}, with $\fmm=0$ (identical to the $\sigma=0$ baseline model), $\fmm=0.3$, and $\fmm=0.6$ (thick black solid, dashed, and dotted curves, respectively), for the first three objects in our sample. As discussed in \S\ref{subsec:effective_fractions}, the effective ions fractions and column densities of high ions are boosted significantly even for small $\fmm$, with a linear increase for higher values. For this model we show the OVI column density, and for motivated values of $\fmm$ this model produces large OVI columns, comparable to observations. The thin solid green curve is the self-shielded model with $\fmm=0$.}
\label{fig:fits_cool}
\end{figure*}

\begin{figure*}
{\centering{\includegraphics[width=1.0\textwidth]{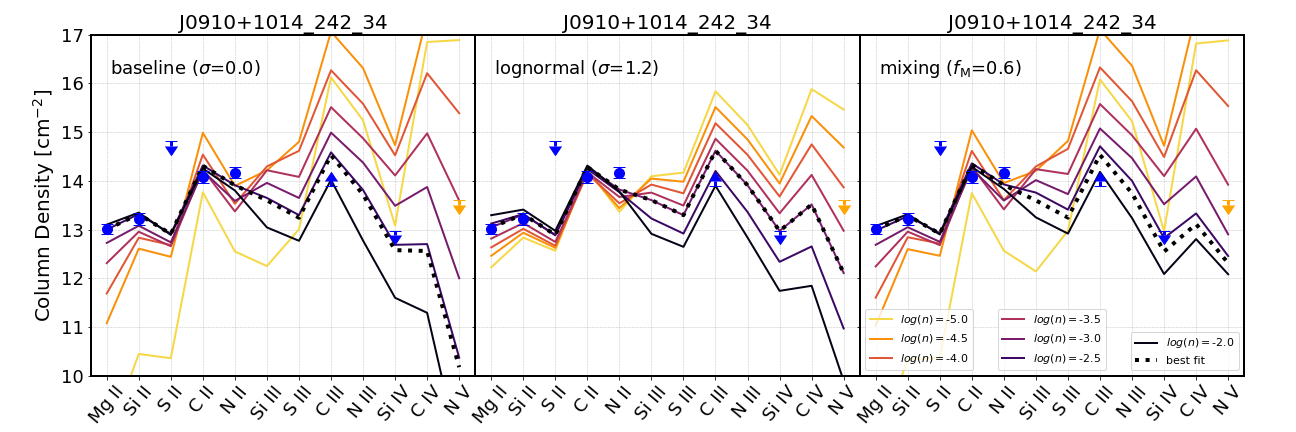}}}
\caption{The effect of gas density on the ionization state and metal ion columns. Similar to Figure~\ref{fig:fits_lognormal} for the first system in our sample, showing how the columns vary with gas density for the baseline model (left), lognormal isothermal scenario with $\sigma=1.2$ (middle panel), and the mixing layers model with $\fmm=0.6$ (right). While the baseline model produces a wide range of columns for the low and high ions, models with a density distribution result in a smaller range (see \S\ref{subsec:fit_results}).}
\label{fig:system_scan}
\end{figure*}

\subsection{Fitting results}
\label{subsec:fit_results}

Our fitting results are summarized in Table~\ref{tab:fit_results} and Figures~\ref{fig:fits_lognormal}-\ref{fig:nh_compare}. Table~\ref{tab:fit_results} lists the system ID, and the number of detections and total constraints (detections and limits) used in the fits, $N_{\rm det}$ and $N_{\rm tot}$, respectively (all from \citetalias{Werk13} and \citetalias{Prochaska17}). The following columns show the best-fit density and $\chi^2$ value for a case with no density distribution (hereafter the baseline model) and for the four models we consider in this work (isothermal and adiabatic turbulence, mixing layers, and cooling gas), for different values of the model parameter ($\sigma$ or $\fmm$). For the sake of compactness, for most models we only present here the results for the maximal value of model parameter ($\sigma=1.2$ or $\fmm=0.6$) and we provide the full results in an attached data file. The last two columns show the fitting results for the self-shielded scenario, in which the MGRF intensity is set to $0$ at energies above $4$~Ryd (see \S\ref{subsec:HeII_SS} for details), with no density fluctuations and $\sigma=1.2$. In the figures below we plot results for the lognormal isothermal and the mixing layers models, since as we showed in \S\ref{sec:scenarios}, the adiabatic case is less likely to occur, and the results for the cooling gas model are similar to those of mixing layers. Figure~\ref{fig:fits_lognormal} shows the observed column densities and the model columns for the isothermal lognormal scenario, where the thick black curves are the best case fits for optically thin gas with model parameters of $\sigma = 0, 0.6,$ and $1.2$ (solid, dashed, and dotted, respectively, identical to the styling in Figures~\ref{fig:eff_frac1}-\ref{fig:eff_frac2}). The thin solid green curves show the fits for the \HeII\ self-shielded case, for $\sigma=0$. The text at the top of each panel indicates the best fit $n_0$ and $\chi^2$ values for these four models, in the order described above. Figures~\ref{fig:fits_mix} and \ref{fig:fits_cool} show the same for the mixing layers and cooling gas scenarios, respectively, for the first three systems in our list, for brevity. The ions in these figures are ordered by increasing ionization energy (needed to form the ion), from $E=7.6$~eV for \MgII~to $113.9$~eV for \OVI\ (see Appendix~D in \citealp{Draine11ISM}).

First, we examine the goodness of fit of the models to the measurements. Figure~\ref{fig:fits_lognormal} shows that for most objects, the best fit model columns for the optically thin case (plotted by the thick black curves) are in reasonable agreement with the reported measurements and limits (shown by the blue markers). For a more quantitative look, the $\chi^2$ values are listed at the top of each panel and in Table~\ref{tab:fit_results}. With two free parameters being fit, $\chi^2 \sim (N_{\rm det} - 2)$ would indicate a good fit if the upper and lower limits are not important. However, many of the limits do play a role in constraining the models so that a good fit is better indicated by $\chi^2 \sim (N_{\rm tot} - 2)$, where $N_{\rm tot} \equiv N_{\rm det} + N_{\rm lim}$. Since $N_{\rm tot} \approx 6-10$, we take $\chi^2 \sim 10$ to indicate a decent fit. Note that there is some uncertainty in the size of the error bar which is set to a floor value of $\sigma_{N}=0.1$, due to observational and theoretical uncertainties discussed earlier, except for a few measurements for which the observational uncertainty reported in \citetalias{Werk13} is larger.  Thus, we assess a good fit as one in the ballpark of $\chi^2 \sim 10$. Seven out of the nine objects are consistent with this threshold for at least some of the models and parameters. We discuss the two remaining objects in detail at the end of this section.

Our primary interest and goal in this work is to determine whether a broader PDF of densities and temperatures leads to an improved fit. A change of $\Delta \chi^2 = 4-9$ would formally indicate a $2-3\sigma$ preference for the model that returns the lower $\chi^2$. For the isothermal turbulence scenario, $\chi^2$ only decreases significantly for two of the well-fit objects (and for $\sigma=1.2$), which are $274\_6$, with $\Delta \chi^2=8.3$ and $197\_23$, with $\Delta \chi^2=4.2$. Only for the first object one could argue that the fluctuations are really significantly preferred statistically, and we discuss it in detail at the end of this section. For the two worst-fit systems ($208\_27$ and $289\_28$), $\chi^2$ does decrease significantly with $\sigma$ (with $\Delta \chi^2 \approx 20$ for $\sigma=1.2$), preferring a wider distribution, but is still insufficient to yield a good fit. For the remaining five objects $\chi^2$ only changes mildly with $\sigma$ (either decreases or increases). We conclude that while lognormal density distributions can produce column density variations, potentially allowing absorption data to differentiate between models with and without fluctuations, for the objects in our sample such distributions are not preferred relative to the baseline model. For the adiabatic scenario, all the objects have higher $\chi^2$ values than the isothermal model, ranging from small to significant $\Delta \chi^2$.

The data are even less constraining for the mixing and cooling scenarios, as for these models the distributions have a smaller effect on the total columns of the low and intermediate ions (see Figures~\ref{fig:fits_mix} and \ref{fig:fits_cool}). In these scenarios, none of the objects show a significant improvement with larger or smaller $\fmm$. One explanation is that at the gas densities inferred from the data ($\nh \lesssim 10^{-3}~\cmv$), these models have a significant effect only on the higher ionization states (\CIV, and \NV, for example, as shown in Figure~\ref{fig:eff_frac2}), and the data set we use to constrain the models only includes very few measurements of these ions, if at all. Upcoming observations that include higher ions may be able to provide better constraints on models with gas at intermediate temperatures (see \S\ref{subsec:predictions}).

A caveat to the conclusion that the mixing and cooling models are less constrained is if \OVI\ is included in the fits rather than assuming that it owes to unmodeled gas, such as virialized gas. In this case, the cooling gas model (Figure~\ref{fig:fits_cool}) does produce significant ion fractions, and column densities that are comparable to existing measurements. For some of the systems, the measured columns can even constrain the amount of mixing and cooling gas, ruling out large values of $\fmm$. Another way to constrain the value of $\fmm$ is using the \OVI/\NIII\ ratio. For example, in $\rm 94\_38$ $N_{\rm \text{\OVI}}/N_{\rm \text{\NIII}} \approx 1$, and our models suggest $\fmm \lesssim 0.2$, similar to the estimate from Eq.~\eqref{eqn:cool_est2}.

Finally, how does including density distributions in models affect our inferences of the gas properties? The $x-$axii in Figure~\ref{fig:nh_compare} shows our density and metallicity best fit values, including their uncertainties, indicated by the error bars. The $y$-axis plots the best-fit values of \citetalias{Prochaska17}, and we compare these in \S\ref{subsec:comparison}. The baseline model is plotted by the black squares, the isothermal turbulence scenario is shown by the blue and green circles, and the mixing layer model is shown by the red diamonds. The gas densities are plotted in the left panel, and for the isothermal scenario the mean densities are lower in the presence of density distributions. The values inferred for the $\sigma=0.6$ case (small blue circles) do not differ from those of the baseline model for about half the objects and are consistent with them within $\approx 2\sigma$ for the other half. For the widest lognormal distribution, with $\sigma=1.2$ (large green circles), the inferred densities are lower than those of the baseline model by factors of 3-10, differing from the baseline model by more than $2 \sigma$ for most of the objects. For the mixing layers scenario, the plotted densities are those of the cool photoionized component, and they are close to to the densities inferred for the baseline model, including being offset by less than $1\sigma$.\footnote{We also note that repeating the analysis using $\teq$ for instead $Z=1$ has a small effect on the inferred gas densities, at the level of $\approx 0.1$ dex, which adds a systematic, model uncertainty on the inferred densities.}

The $x-$axis in the right panel shows our inferred gas metallicities, and for these the differences between the different models are small, both in absolute values and compared to their uncertainties. All of the metallicities inferred in the presence of density distributions are consistent with the baseline model values within $1\sigma$.

Other than the gas densities and metallicities, we also infer the total hydrogen column, $\NH$, and the absorption pathlength, $L$. We provide these results in the data files attached to the paper. The total columns are between $10^{18}$ and $6 \times 10^{19}~\cmc$ for 8/9 systems, and a single system with $2 \times 10^{17}~\cmc$. The results for $\NH$ do not vary significantly with the model parameter or between scenarios - for 7/9 objects in the sample, the difference from the baseline model is less than $\approx 35\%$. This suggests that while the gas volume density can be sensitive to the presence of fluctuations, the total amount of gas, given by the product of the gas density and total pathlength, is a more robust result. Given the lack of direct independent constraints on cloud sizes in absorption observations, the gas pathlength can be estimated as the total column divided by the gas density. For the models and systems we examine, the pathlengths are in the range $0.3 < L/{\rm kpc} < 30$, with a typical pathlength of $\approx 1-3$~kpc. Since for isothermal turbulence large $\sigma$ leads to low densities, these models have larger $L$. Spatially resolved emission observations can provide better constraints on the gas morphology, both through an estimate of the gas extent on the sky (albeit perpendicular to absorption pathlength, so still requiring some assumptions), and through joint modeling of emission and absorption data (see \citealp{2016ApJ...827..148C, Johnson22, 2022MNRAS.516.3049P, 2023arXiv231100856N}, and \citealp{Aspera21, Maratus23} for upcoming missions).

Finally, to better understand our results, we examine how the gas density affects its ionization state and metal ion columns. Figure~\ref{fig:system_scan} shows the variation in ion columns across the range of densities we consider in our analysis for a single system for three models -- baseline (left panel), lognormal isothermal with $\sigma=1.2$ (middle), and mixing layers with $\fmm = 0.6$ (right). Similar to Figures~\ref{fig:fits_lognormal} and \ref{fig:fits_mix}, the markers show the measured columns, while the solid curves show the model outputs, in this case for densities between $10^{-5}$ (yellow) and $10^{-2}~\cmv$ (black), in $0.5$~dex steps. The thick dotted curve shows the best fit density for this system, which varies between models. The figure highlights two phenomena related to the gas ionization state. First, the ions on the x-axii are ordered by ionization energy, and the gas becomes less ionized with increasing density, with the columns of low ions (\MgII, \SiII, and others) increasing with density, while the columns of high ions (\CIV, \NV) decreasing. Second, and more important for this work, the range of possible columns of a given ion is smaller for models with density distributions, a reflection of the broadening of the ion fraction curves as functions of the density seen earlier in Figure~\ref{fig:eff_frac1}. For example, for the range of densities shown here, \MgII\ column densities span more than three dex ($\approx 10^{10}-10^{13}~\cmc$) in the baseline model, while in the lognormal model, this range is reduced to less than one dex. Similarly, the range of \SiIV\ columns is reduced from three to about two dex.

\begin{figure*}
{\centering{
\includegraphics[width=0.49\textwidth]{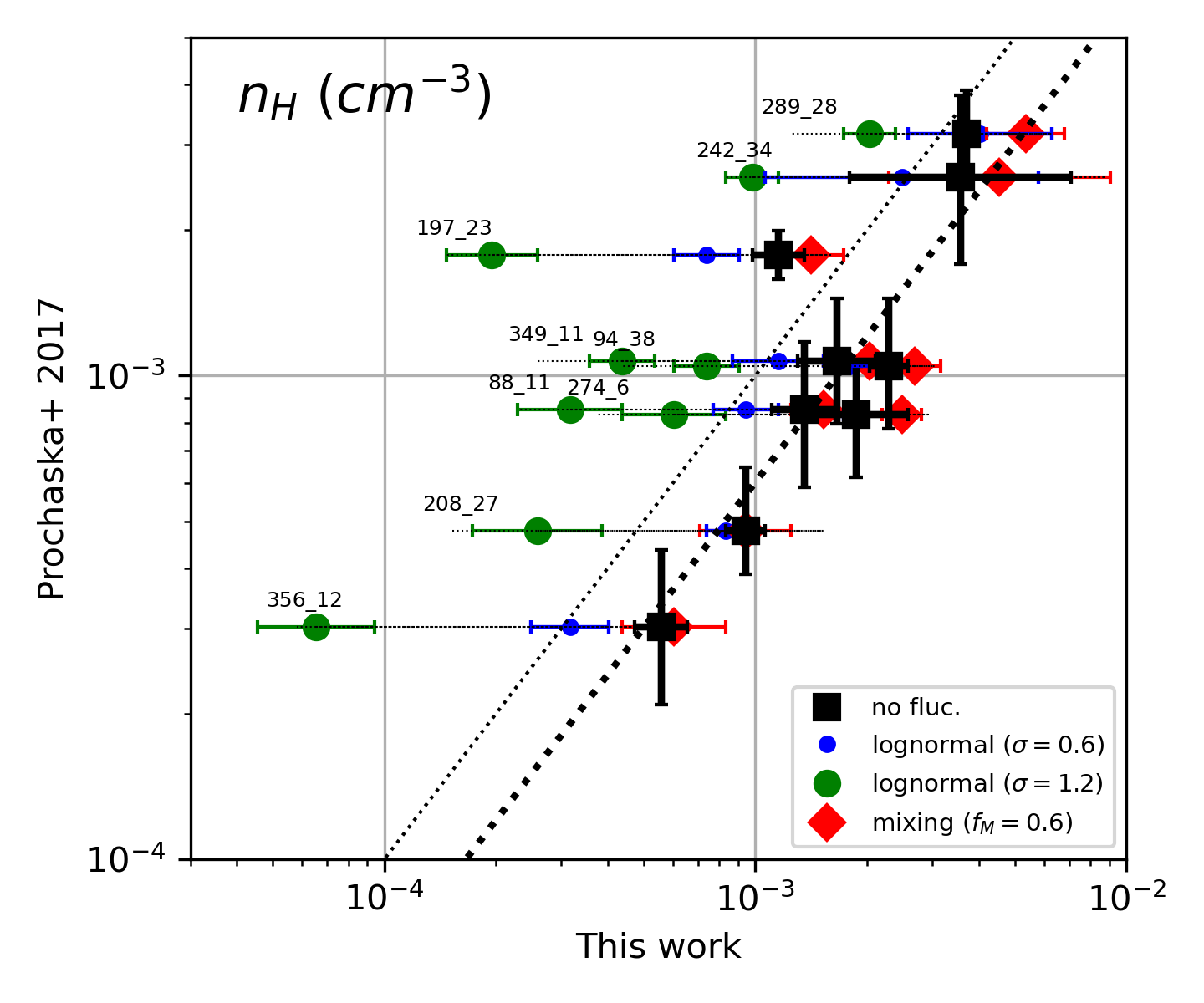}
\includegraphics[width=0.49\textwidth]{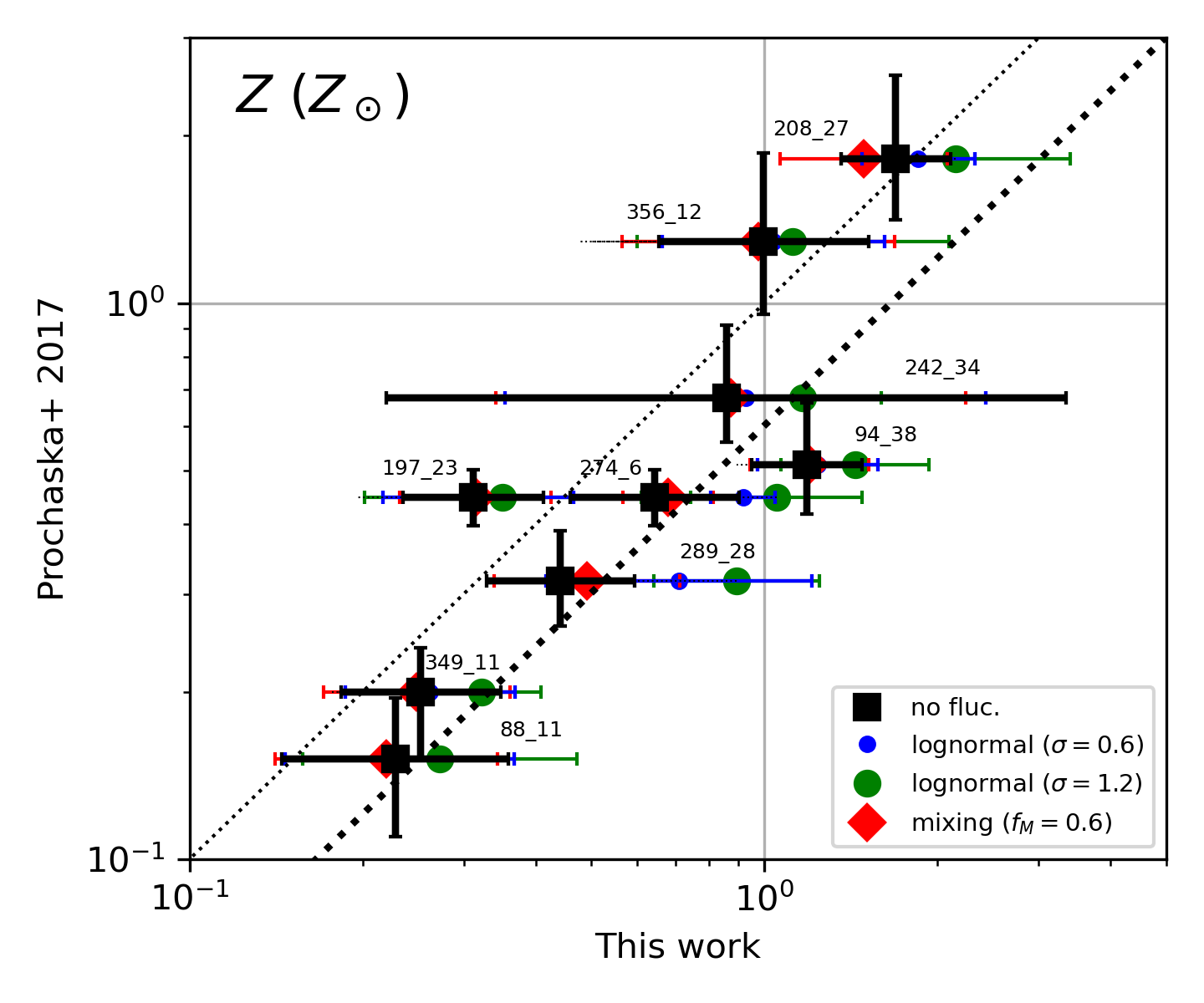}}}
\vspace{-0.3cm}
\caption{Comparison of our results for the gas densities (left) and metalllicities (right) to the values inferred by \citetalias{Prochaska17} (text tags show object IDs). Black squares show the results for the baseline model, without density fluctuations, and these are consistent with the results in \citetalias{Prochaska17} when adjusted for the different MGRF used in this work (thick dotted line). Circles show the densities inferred in the presence of lognormal isothermal distributions with $\sigma=0.6$ and $\sigma=1.2$ (small blue and large green markers, respectively), and large values of $\sigma$ lead to significantly lower densities and slightly higher metallicities. Red diamonds show the results for the mixing layers model with $\fmm=0.6$, which has a smaller effect on the inferred gas properties compared to the baseline model. The gas properties inferred for the cooling gas model are nearly identical to those of the mixing layers model and are not plotted here (see Table~\ref{tab:fit_results} and \S\ref{subsec:comparison}).}
\label{fig:nh_compare}
\end{figure*}

\subsection{\HeII\ self-shielding}
\label{subsec:HeII_SS}

Photoionization modeling of objects near the threshold column density where hydrogen or helium can self-shield has additional uncertainty as the ionizing background within the system depends on its unknown geometry. While we avoid objects with \HI\ self-shielding in our sample selection, \HeII, absorbing photons at  $>4~$Ry, starts to shield at a lower \HI\ column density and may still be relevant in some of the systems, potentially affecting the ionization state of the higher metal ions (\CIV\ and \SiIV). We now estimate when shielding by \HeII\ starts to occur and check how it impacts our modeling inferences. In Appendix~\ref{app_HeII} we also examine in detail the effect of \HeII\ self-shielding on individual metal ion fractions.

The critical \HeII\ column for self-shielding is given by $N_{\rm \rm HeII}^{\rm crit} = \sigma_{\rm HeII}^{-1} = 6 \times 10^{17}~\cmc$, where $\sigma_{\rm HeII}$ is the \HeII\ photoionization cross section at $4~$Ry. Above this column density, the $4$~Ry background would be strongly attenuated as it penetrates the cloud. We can write the \HI\ column density that corresponds to $N_{\rm HeII}^{\rm crit}$ as
\begin{equation}\label{eqn:nhi_crit}
N_{\rm HI}^{\rm crit} = \frac{f_{\rm HI}}{f_{\rm HeII}} \frac{1}{A_{\rm He} \sigma_{\rm HeII}} \approx 2 \times 10^{16} \left( \frac{f_{\rm HI}/f_{\rm HeII}}{0.003} \right)~\cmc ~~~,
\end{equation}
where we used $A_{\rm He} \approx 1/12$, the primordial helium abundance. At IGM and warm/hot CGM gas densities, $\nh \lesssim 10^{-4}~\cmv$, both ion fractions are small and $f_{\rm HI}/f_{\rm HeII} \approx 0.003$ across a broad range of redshifts owing to the spectrum of the MGRF (\citealp{2009ApJ...703.1416F}, \citealp{HM12}, and see also Appendix~A in \citealp{mcquinn09}). This ratio is also independent of density and (essentially) temperature for photoionized gas. At higher densities, the \HI\ ion fraction increases with density faster than that of \HeII, and $N_{\rm HI}^{\rm crit}$ is no longer constant. In Figure~\ref{fig:nhi_crit} we plot the ratio of \HI/\HeII\ ion fractions (right y-axis), and the corresponding value of the \HI\ critical column (left y-axis) as functions of $\nh$. The critical columns  are between $2 \times 10^{16}$ and $4 \times 10^{17}~\cmc$ (shown for three metalicities via the three curves, but the metalicity dependence is extremely weak). 

Objects with \HI\ column densities above the critical value may be self-shielded from $4$~Ry radiation, while at lower columns they should be optically thin. We use the \HI\ columns measured by \citetalias{Prochaska17} and the densities inferred in this work (see \S\ref{subsec:fit_method}) to test whether our assumptions are self-consistent, and the results are plotted by the markers in Figure~\ref{fig:nhi_crit}. First, assuming optically thin gas (full markers), six out of the nine objects in our sample are below the $N_{\rm HI}^{\rm crit}$ curve and consistent with our assumption, but only three are relatively far from the boundary so that we are confident self-shielding is not important. The remaining three objects, $274\_6$, $349\_11$, and $88\_11$, have high enough \HI\ columns at the inferred density for the \HeII\ to self-shield, although $274\_6$ and $349\_11$ are not far over the threshold.  

To assess the effect of \HeII\ self-shielding on our models, we consider the extreme scenario in which $>4~$Ry photons are fully absorbed by \HeII. We compute the ion fractions with the MGRF intensity set to zero at higher energies using \textsf{Cloudy}, and in Appendix~\ref{app_HeII} we show how this affects the metal ion fractions. We then fit the observational data using the effective ion fractions calculated for the fully self-shielded case with the lognormal (isothermal) distribution. Our fitting results are given in the last two columns of Table~\ref{tab:fit_results} and shown in Figure~\ref{fig:nhi_crit} by the empty markers for $\sigma=0$. The densities inferred with fully self-shielded gas for all objects are lower than for the optically thin regime. This reflects the fact that at a fixed density self-shielded gas is less ionized than optically thin gas, and a lower density (higher ionization parameter) is needed to reproduce the same observed ionization state.

For two out of the three objects above the $N_{\rm HI}^{\rm crit}$ curve (349\_11, and 88\_11) curiously the self-shielded model is not preferred by the data (see Table~\ref{tab:fit_results}). For example, for the case with $\sigma=0$, for these two object the self-shielded model results in $\chi^2=7.4$ and $46.1$, respectively, while the optically thin case gives $\chi^2=2.7$ and $11.8$. One possible explanation is non-spherical absorber geometry -- for an elongated cloud, the column along the major axis (possibly coinciding with our line of sight) will be high, while the column from along the minor axis (and most other angles) will be lower and lead to a lower attenuation of the MGRF. Only for two objects in our sample the self-shielded model has a lower $\chi^2$ over the optically thin case. For $274\_6$, with $N_{\rm HI} > N_{\rm HI}^{\rm crit}$, the self-shielding model gives with $\chi^2=0.3$ compared to $16.3$ for the optically thin case. For $289\_28$, near the $N_{\rm HI}^{\rm crit}$ curve, the two models give $\chi^2 = 40.1$ and $59.1$, respectively. For the latter system, even the self-shielded model does not provide a good fit, and we discuss both objects below. These two results - that absorbers are somewhat below the critical \HI\ column for \HeII\ self-shielding and that the data does not prefer the fully-shielded models - provide justification for our focus on the optically thin regime.

To summarize, (\HeII) self-shielding introduces additional uncertainty in the ionization fractions that potentially depends on the geometry of the absorber in 3D space. Self-shielding can obscure the effects of density distributions, since the two have opposite effects on the fractions and columns of high ions.

\begin{figure}
{\centering{\includegraphics[width=0.49\textwidth]{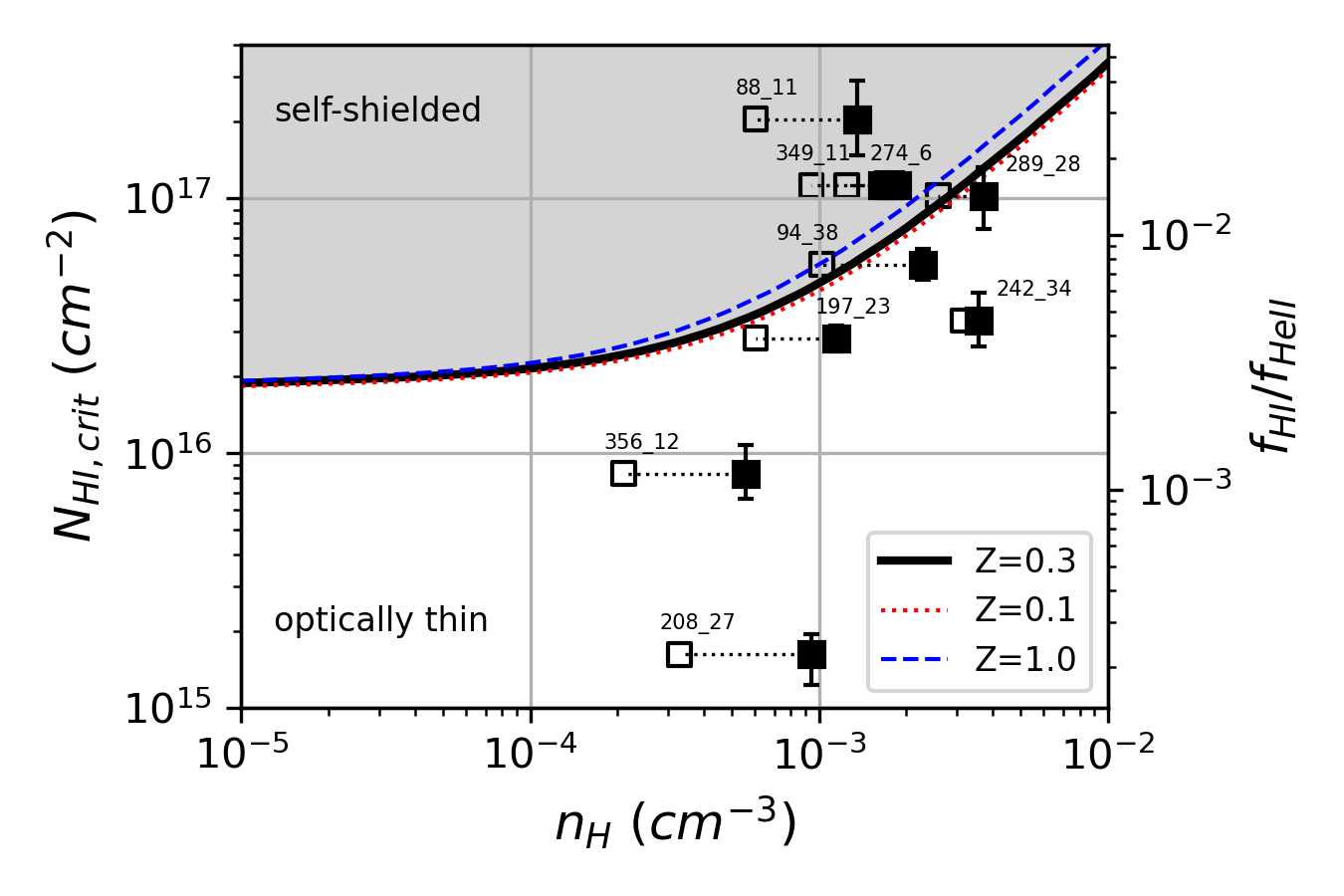}}}
\vspace{-0.5cm}
\caption{The effect of \HeII\ self-shielding in our analysis. The curves show the \HI\ to \HeII\ ion fraction ratio (right axis) and the corresponding \HI\ column that we estimate should self-shield with an optical depth of unity at $4$~Ry (left axis). The black solid curve is this critical column for a nominal metallicity of $Z=0.3$, and the red dotted and blue dashed curves show $Z=0.1$ and $1.0$ solar, respectively. The markers show the best-fit gas densities for the nine COS-Halos absorption systems, assuming optically thin gas (filled) and full \HeII\ self-shielding (empty) versus the measured \HI\ column density (from \citetalias{Prochaska17}). Systems that fall below the curves are likely to be optically thin to $4$~Ry photons, the limit of much of our analysis (see Table~\ref{tab:fit_results} and \S\ref{subsec:HeII_SS}).}
\label{fig:nhi_crit}
\end{figure}

\subsection{Individual objects}
\label{subsec:fit_objects}

We now discuss some of the objects in more detail - one for which fluctuations significantly improve the fit, two which are not fit well with our models (with or without fluctuations), and one for which the \NV\ column density limit actually constrains the models.

\begin{itemize}

\item {\it J1016+4706\_274\_6} ($\chi^2 = 16.3$ for the baseline model): This object is the only one in which the fit is significantly improved in the presence of lognormal density fluctuations, with $\sigma=1.2$ resulting in $\chi^2=8.0$, compared to $\chi^2=16.3$ for $\sigma=0$. This fit also prefers a lower density, with $\nh = 6.0 \times 10^{-4}~\cmv$, compared to a density of $1.9 \times 10^{-3}~\cmv$ for $\sigma=0$, suggesting a preference for higher ionization. However, the $\chi^2$ is even lower for the case of self-shielded gas and no density distribution, with $\chi^2=0.3$ (and $\nh = 1.2 \times 10^{-3}~\cmv$, similar to the optically thin result). The high \HI\ column density measured for this object supports the scenario that the \HeII\ does self-shield (see Figure~\ref{fig:nhi_crit}). Adding density fluctuations in the self-shielded case does not reduce the $\chi^2$ further, but does decrease the gas density to $3.7 \times 10^{-4}~\cmv$.

\item {\it J1241+5721\_208\_27} ($\chi^2=84.7$): The most significant discrepancy between the models and the measurements is coming from the Silicon ions. The measured Silicon columns are all similar, which is inconsistent with a single density. Specifically, the large measured \SiIV\ column is not reproduced by the models that fit the lower ions, even with the low gas densities producing the best fit -- $\nh = 9.4 \times 10^{-4}~\cmv$ for $\sigma=0$, and $\nh = 2.6 \times 10^{-4}~\cmv$ for $\sigma=1.2$, among the lowest densities in our results. Overall, the increased $\sigma$ yields more high ions that are able to better fit especially the observed \SiIV\ column. We note that the HI column density in this object is very low, with $N_{\rm HI} = 10^{15.2}~\cmc$, leading to a high inferred metallicity, of $Z \approx 1.5-2.0$~solar. Possible reasons for the bad fit may be (i) the high metallicity leading to low $\teq$ (see Figure~\ref{fig:temp_eq}) (see Appendix~\ref{app_met_var} for the effect of metallicity on the ion fractions), or (ii) a larger amount of lower density and higher ionization gas, which could result from non-thermal pressure or perhaps different gas pressures contributing to the absorption system  (see \citetalias{FW23}; the latter might be likely if gas at different radii from the galaxy contribute).

Finally, this is the only object in our sample (and one of a few in the COS-Halos full sample) with an actual detection of \NV\ absorption, with $N_{\rm \NV} = 4.3 \times 10^{13}~\cmc$. Our models are not capable of reproducing this column density even for the widest lognormal distribution or the most massive mixing layer, and produce $\approx 2.5 \times 10^{12}~\cmc$ for the lognormal $\sigma=1.2$ model and $\approx 1.5 \times 10^{12}~\cmc$ for the  mixing layer $\fmm=0.6$ model. Including the \NV\ measurement in the fitting process leads to a worse fit, with $\chi^2 \approx 200$.

\item {\it J1330+2813\_289\_28} ($\chi^2=59.1$): Here the model columns of \NIII\ and \SiIII\ are inconsistent with the data, underpredicting and overpredicting the measurements, respectively. Furthermore, the \MgII\ column is below the reported limit. The latter can be explained by cool gas that is not modeled in this work. This object has the highest density value in our sample, both for the baseline model, with $\nh = 3.7 \times 10^{-3}~\cmv$, and in the presence of (lognormal) fluctuations, with $\nh = 2.0 \times 10^{-3}~\cmv$ for $\sigma=1.2$. This baseline density is only marginally above the density inferred for $242\_34$, which has a low $\chi^2$ with $\nh = 3.6 \times 10^{-3}~\cmv$~for the baseline model. However, one potentially interesting difference is that while for the latter object the fit becomes (slightly) worse for the lognormal model ($\Delta \chi^2 = 3.2$), for $289\_28$ it improves significantly ($\Delta \chi^2 \approx 20$), potentially suggesting that an even wider distribution could further improve the fit.

\item {\it J2345+0059\_356\_12} ($\chi^2=3.2$): This is the only object for which the reported \NV\ column density limit provides a constraint on the lognormal models. Indeed, the best-fit $\sigma=1.2$ model has $\chi^2=4.9$ but it is inconsistent with the reported \NV. Using this limit in our fitting gives $\chi^2 = 12.2$ due to a \SiIV\ column density below the measured value.

\end{itemize}

\section{Discussion}\label{sec:discussion}

\subsection{Comparison to previous works}\label{subsec:comparison}

For the COS-Halos data set, \citetalias{Werk14} used photoionization modeling of low and intermediate metal ions column densities to infer the gas densities along the line of sight. \citetalias{Prochaska17} combined updated measurements of \HI\ column densities with the metal absorption reported by \citetalias{Werk13} to better constrain the metallicities for some of the lines of sight. Since \citetalias{Werk13} and \citetalias{Prochaska17} modeled the ionization with the \textsf{Cloudy} code, \HeII\ (and \HI) self-shielding is accounted for by assuming a slab geometry. In Figure~\ref{fig:nh_compare}, we compare our results for the baseline model (black squares) to the results reported by \citetalias{Prochaska17} for the gas densities and metallicities and their uncertainties.

For the gas densities (left panel), the offset between the two studies is caused by the different radiation fields used in each work --  \citetalias{Prochaska17} use the \citet{HM12} MGRF, whereas we adopt the \citetalias{KS19} fiducial model, which has an ionizing flux higher by a factor of $1.66$. The thick dotted line shows the one-to-one density relation with this factor accounted for, and our results cluster around this line. For the metallicity (right panel), the ratio of values inferred by the two works is close to unity. Our metallicities have a slight systematic offset to higher values, which may be the result of the different radiation field we adopt or the way limits are used in our analysis. However, the individual inferred values are consistent with the one-to-one relation within their uncertainties.

As noted earlier, we also infer the total hydrogen column, $\NH$, and the absorption total pathlength, $L$, and provide these in the data files attached to the paper. Our results for $\NH$ agree well with the column densities inferred by \citetalias{Prochaska17} (private communication), suggesting that the total column density is fairly robust to modeling methods and assumptions. The path lengths we infer, in the range of $0.3-30$~kpc, are similar to the estimates by \cite{Stocke13} and \cite{Werk14} and consistent with the product of the cloud size and number of clouds in \citetalias{FW23} (see their Figure 11).

A different method of extracting the gas properties from absorption data uses component fitting (see \citep{Sameer21,Qu24} and earlier references). In this approach, the spectra are fitted with multiple (Voigt) line  profiles, interpreting each as corresponding to an physical gas cloud. This method is, by construction, discrete, assuming the existence of some number (a free parameter of the analysis) of individual absorbers with varying physical conditions which may or may not be related to one another in position or thermal properties. Our approach, in contrast, assumes a continuous distribution of gas densities and temperatures, with the distribution shapes and relations between gas properties set by a physical scenario (turbulence, mixing, cooling, isothermal, adiabatic, isobaric, etc.). 

\subsection{Redshift scaling}\label{subsec:redshift}

In this work we calculate the ion fractions in the presence of the \citetalias{KS19} MGRF at $z=0.2$, the typical redshift of the COS-Halos galaxies. Previous studies such as \citetalias{Prochaska17} have taken the same approach, and we report our results at a single redshift for ease of comparison (see \S\ref{subsec:comparison}). This results in the true density and metallicity of the absorber being (slightly) different than the inferred values, and we now estimate this effect.

In PIE and to the extent that the shape of the ionizing background is constant (which is likely a good approximation at $z<2$, see \citealp{2009ApJ...703.1416F,HM12}), the gas ionization state is determined by the ionization parameter, defined as $U \equiv \Phi/(c \nh)$, where $\Phi$ is the ionizing photon flux at energies above $1$~Rydberg and $c$ is the speed of light. Thus $\nh \propto \Phi$, allowing us to scale our results for the inferred gas densities to other redshifts. The total ionizing flux can be approximated as a power-law function of redshift, $\Phi \approx \Phi_0 (z+1)^\alpha$, and at low redshifts, it is a strong function of $z$. For example, for the \citetalias{KS19} fiducial model, $\alpha \approx 4.5$ and $\Phi_0 \approx 1.6 \times 10^4~{\rm photons~s^{-1}}$ for $z<0.5$, encompassing the full range of redshifts of the COS-Halos galaxies\footnote{We note that extending the fit of $\Phi(z)$ to $z = 1$ ($z=2$) produces a shallower slope, $\alpha \approx 4.17~(3.15)$, with $\Phi_0 = 1.8 \times 10^{4}~(2.5\times 10^{4})~{\rm photons~s^{-1}}$.}. The objects in our subsample have $0.19<z<0.31$, and for an object at $z=0.3$, the density we infer should be increased by a factor of $\approx 1.4$. Most of the absorbers are at redshifts close to $0.2$ where this correction is negligible. As described in \S\ref{subsec:distributions}, the gas equilibrium temperature also depends on the redshift, with $\teq \propto (1+z)^{0.617}$ (see Eq.~\ref{eqn:teq_fit}). Due to the weaker dependence, this has an even smaller effect on the results for individual galaxies compared to the ionizing flux.

These fits and estimates account for the MGRF, whereas the galaxy may also contribute to the ionization and heating of the CGM. For example, stellar and AGN feedback may inject additional energy into the CGM, either through ionizing radiation, or heating by shocks and other non-thermal mechanisms. \citet{Sanderbeck18} estimated that stellar radiation can affect gas out to $10-30~{\rm kpc} \times ({\rm SFR}/\msuny)^{1/2}$. \citetalias{FW23} use the simulations results from \citet{Sarkar22} to estimate that ionizing radiation from hot outflows should be subdominant to the MGRF at radii $\gtrsim 14$~kpc (or $\gtrsim 0.05~\rvir$ for a MW-mass halo) and ${\rm SFR}=4.3~\msuny$ the median value in the COS-Halos sample (see their Section~6.2). The galaxies in our subsample have ${\rm SFR}<4.4~\msuny$, and the absorbers are at impact parameters of $h \geq 23~{\rm kpc}$ ($h \geq 0.1~\rvir$). Thus, we do not expect them to be affected significantly by galactic radiation.

\subsection{Predictions}\label{subsec:predictions}

Figure~\ref{fig:fits_lognormal} shows the results of our models for the column densities of the entire set of ions measured in the COS-Halos survey. For ions that are not measured or detected for individual objects, these serve as predictions, and we now discuss some of these.

\SII\ and \SIII\ have upper limits for all the objects in our sample, typically at $ 1-3 \times 10^{14}~\cmc$. Our models are consistent with these limits, with the predicted column densities in most objects lower than the limits by at least a factor of $3-10$ for \SIII, and by more than an order of magnitude for \SII. Two exceptions are (i) $274\_6$, where the models predict $N_{\rm SIII} \approx 10^{14}~\cmc$, just at the estimated limit, and (ii) $88\_11$, for which the \SIII\ column is not measured and the models also predict $\approx 10^{14}~\cmc$. Targeted observations of these two lines of sight may be able to provide more constraining limits or even to detect this ion, allowing to test the presence of distributions.

\NV\ is rarely detected in the COS-Halos data set, with most measurements providing upper limits around $3-10 \times 10^{14}~\cmc$, and a few detections with $\sim 6 \times 10^{13}~\cmc$. For most of the objects in our sample, our models predict $N_{\rm \NV} < 3 \times 10^{12}~\cmc$, significantly below the reported limits. However, there are two exceptions -- $88\_11$ and $356\_12$, which for $\sigma=1.2$ lognormal fluctuations have columns of $N_{\rm \NV} = 3-10 \times 10^{13}~\cmc$, around the reported limits and potentially observable. More sensitive \NV\ column measurements in these systems  would be constraining.

\CIV\ absorption is not covered by most of the COS-Halos spectra and is unconstrained in \emph{all} of our sightlines, but is targeted by the $CIViL^*$ survey (PI: T. Berg, \citealp{Berg_civil}). In our models, the \CIV\ column densities are between $3 \times 10^{12}$ and $10^{14}~\cmc$, and are sensitive to the density distribution width, with variation by a factor of $3-10$ between $\sigma=0$ and $1.2$ for about half the objects in our sample. We predict columns of $\approx 1-3 \times 10^{14}~\cmc$ for all values of $\sigma$ in objects $274\_6$, $88\_11$ and $356\_12$, and columns of $\approx 10^{14}~\cmc$ for the high $\sigma$ model in objects $94\_38$, $349\_11$, and $197\_23$. These columns are similar to the few measured \CIV\ columns in the COS-Halos data set, and to detections in other COS data \citep{Bordoloi14,Borthakur15,Burchett16}.

Finally, we note that \NII\ is sensitive to \HeII\ self shielding, with a variation by a factor of $\approx 2-3$ between the thin and self-shielded models in about half the objects. In some of the objects it is detected and provides strong constraints on the models, and targeting it in additional lines of sight ($356\_12$, $197\_23$, $349\_11$) could test our indirect methods for assessing self-shielding.

\section{Conclusions}\label{sec:summary}

In this work, we examined the detectability of physically-motivated density and temperature distributions in the cool CGM by comparing theoretical models with absorption measurements.

First, we presented several scenarios for density and temperature distributions - (i) {\bf lognormal density distributions}, which can result from turbulence in the CGM or alternatively from the superposition of many processes, (ii) {\bf mixing layers} at the boundary of cool clouds, which lead to a uniform distribution in temperature, and (iii) {\bf gas cooling} from the hot phase to the cool, with the probability that gas resides at a given density proportional to the gas cooling time (see \S\ref{sec:scenarios}). By and large, our results show that for physically-motivated values for the model parameters, density and temperatures distributions have a fairly modest effect on the ion fractions compared to a model that assumes single-density photoionized gas. This suggests that in many cases single-density models can be a reasonably good approximation and that the effects of such density and temperature fluctuations may be challenging to disentangle from observations.

Specifically, for the lognormal scenario motivated by turbulence, linewidth constraints suggest that the (natural) logarithmic width of this distribution is $\sigma \lesssim 1.2$. We considered an (approximately) isothermal model, with gas at the MGRF heating/cooling equilibrium temperature, $\teq$. We also considered an adiabatic case ($T \propto n^{2/3}$) that we argued was less physically likely, showing that this scenario can only hold if the driving scale of the turbulence is small. For both cases, we found that at mean densities for which the ion fractions peak in PIE, the effective ion fractions are only weakly affected compared to the case without density fluctuations, decreasing by $\lesssim 30\%$. However, at mean gas densities lower than this value, the effective ion fractions can increase significantly, by factors of up to $\sim 10$ because as the distribution samples densities that have much higher ion fractions (see Figure~\ref{fig:eff_frac1}). For the adiabatic scenario, the effective fractions of the higher ionization-state ions can also be boosted at higher densities, due to the higher temperatures.

For the cooling and mixing scenarios, we parameterize the distribution using the ratio of masses in intermediate to cool gas phases, $\fmm$, and use that the boundary layer width must be less than the cloud size and that the amount of intermediate temperature gas cannot overproduce the \OVI\ columns to bound $0 \leq \fmm \lesssim 0.6$. For these models we consider isobaric gas, and the effect of the distributions is similar in the two scenarios, with an increase in the ion fraction at densities above the peak density in PIE, especially for the intermediate ions like \SiIV, \CIV\ and \NV\ (see Figure~\ref{fig:eff_frac2}).  The cooling scenario produces more of these intermediate ions, even enough to explain the observed \OVI\ columns.

Second, we compared the outputs of our models to the column densities measured in absorption studies. We used the COS-Halos survey data, selecting lines of sight with (i) more than 3 ions detected per system, to avoid overfitting, and (ii) $N_{\rm HI} < 10^{17.5}~\cmc$, to avoid \HI~self-shielding. Our models provide good fits for most of the systems in our sample. However, for most of the systems and in all of the scenarios we consider in this work, models with density distributions do not significantly improve the fits compared to single density and temperature models (see Table~\ref{tab:fit_results} and Figures~\ref{fig:fits_lognormal}-\ref{fig:fits_mix}).
Thus, the systems we analyzed are largely not able to confirm or rule out variations models that have significant density fluctuations or motivated amounts of cooling and mixing gas, although we note that the constraints are not far from being able to do so. This is particularly true for the lognormal distributions we considered. Additionally, in the lognormal isothermal model, large density distributions do lead to lower inferred mean densities relative to a model with no fluctuations despite both generally providing a good fit to the data, with the $\sigma=1.2$ model resulting in densities that are factor of $3-10$ smaller.

We highlighted that \HeII\ self-shielding may be a significant uncertainty in searching for small effects in ionic columns. Whether an absorber self-shields depends on the absorber size and geometery and gas density, and we showed self-shielding can have a significant effect on the metal column densities, reducing the columns of high ions. Our estimates suggest that 6/9 of the systems in the COS-Halos sample that we fit are \HeII-optically-thin, although some just barely so (see Figure~\ref{fig:nhi_crit} and \S\ref{subsec:HeII_SS}). We tested whether a model with full \HeII\ self-shielding was preferred in all systems, finding that it generally does not provide a better fit (see Table~\ref{tab:fit_results}). Overall, self-shielding may obscure the effect of density fluctuations, with the two having opposite effects on the high ion columns.

Finally, for ions that are unobserved or have upper limits in the COS-Halos data set, our model column densities serve as predictions. For some of these ions, such as \CIV\ and \NV, density distributions do produce a variation in the ion fractions and column densities with the model parameter (see Figure~\ref{fig:fits_lognormal}). Deeper observations of these systems may allow us to better constrain the presence (or lack) of density and temperature distributions. Such constraints on the presence of density fluctuations constrain the circumgalactic medium properties and behavior, offering insights into the flows of gas and energy into and out of galaxies, and improving our understanding of the evolution of galactic ecosystems.

\section*{Acknowledgements}

We would like to thank Erik Solhaug and Kirill Tchernyshyov for helpful comments and discussion during the course of this work. This work is supported by the NASA award 19-ATP19-0023 and NSF award AST-2007012. JKW additionally acknowledges support from NSF-CAREER 2044303.

\software{
{\tt matplotlib} \citep{Hunter:2007},
{\tt numpy} \citep{harris2020array},
{\tt pandas} \citep{pandas_2024},
{\tt scipy} \citep{2020SciPy-NMeth}.}



\appendix
\restartappendixnumbering

\section{Temperature and Density Distributions}
\label{app_dist}

In \S\ref{sec:methods} we presented the effective ion fractions we calculated in the presence of different density distributions. We now describe in detail our derivation of the distribution functions for the mixing layers and cooling gas physical scenarios.

For {\bf mixing layers}, we start with a constant probability as a function of temperature, $p(T) = {\rm const.} \equiv A$. We assume an isobaric layer, so $nT = n_0 \teq = {\rm const.} \equiv P_0/\kb$, where $n_0$ is the cool gas density and $\teq$ is the equilibrium temperature at that density. The distribution as a function of the gas density can be obtained:
\begin{equation}\label{eqn:mix_init}
g(n) = p(T)\frac{dT}{dn} = A\frac{P_0/\kb}{n^2}.
\end{equation}
Rather than normalizing the probability distribution to unity we will normalize so that $\int n P(T) dT = \int n g(n) dn$ is the mass in the mixing layer, and we will follow the same convention for cooling gas.

To calculate the normalization factor, $A$, we can write the ratio of mixing and cool gas masses as
\begin{equation}
\begin{split}
M_{\rm mixing} &~ = \fmm M_{\rm cool} = \int{n g(n) dn} = \\ 
        &~ = A \frac{P_0}{\kb} \ln{\left(\frac{n_0}{n_{\rm min}}\right)} = 
                A \frac{P_0}{\kb} \ln{\left(\frac{\thot}{\teq}\right)}.
\end{split}
\end{equation}
We normalize the cool gas mass to unity, and $A$ is given by
\begin{equation}
A  = \frac{\fmm \kb/P_0}{\ln{\left(\thot/\teq\right)}} ~~~,
\end{equation}
where $P_0$ is the cool gas pressure, and $\thot$ is the (assumed) temperature of the hot phase. Thus, for a given mass ratio and the cool gas equilibrium temperature, the normalization of the PDF varies only weakly with $\thot$. Inserting this into Eq.~\eqref{eqn:mix_init} we get
\begin{equation}
p(T) = \frac{\fmm \kb/P_0}{\ln{\left(\thot/\teq\right)}} ~,~
g(n) = \frac{\fmm}{\ln{\left(\thot/\teq\right)}} \frac{1}{n^2} \\
\end{equation}

For {\bf cooling gas}, we can write down the mass flux through a temperature element $dT$ (or density element $dn$) as
\begin{equation}\label{eqn:coolflux}
\dot{M} = n p(T)\frac{dT}{dt} = n g(n)\frac{dn}{dT} \frac{dT}{dt}.
\end{equation}
We can obtain the time derivative from the 1st law of thermodynamics:
\begin{equation}
\frac{d}{dt}\left(\frac{3}{2}n \kb T V \right) + n k T \frac{dV}{dt} = V n^2\Lambda.
\end{equation}
Since we assume isobaric gas, $P = n\kb T = {\rm const.}$, we can write
\begin{equation}
\frac{5}{2} P \frac{d\ln V}{dt} =  n^2\Lambda ~~ \rightarrow ~~ \frac{d\ln T}{dt} = \frac{2n^2\Lambda}{5P} = \frac{2n\Lambda}{5\kb T},
\end{equation}
where we also used the fact that $V \propto n^{-1}$ such that isobaric gas $V \propto T$. Inserting this into Eq.~\eqref{eqn:coolflux} gives
\begin{equation}
p(T) = \frac{5\kb^3 T^2 \dot{M}}{2 P_0^2 \Lambda} ~~,~~
g(n) = \frac{5P_0\dot{M}}{2n^4 \Lambda}.
\end{equation}
where we used again the isobaric cooling assumption. Inserting the expression for the gas pressure gives:
\begin{equation}
p(T) = \frac{5\kb \dot{M}}{2 n_0^2 \Lambda} \left(\frac{T}{\teq} \right)^2 ~~,~~
g(n) = \frac{5\kb n_0 \teq \dot{M}}{2n^4 \Lambda}.
\end{equation}
To change the normalization from $\dot{M}$ to $f_M$, we use that $f_M  = \int dn n g(n)/M_{\rm cool}$, which allows us to solve for $f_M$ in terms of $\dot M$ and $M_{\rm cool}$. Empirically motivated values for both  $\dot M$ and $M_{\rm cool}$ are discussed in \S\ref{subsec:int_cooling}.

Another approach to normalizing the cooling gas PDF is writing
\begin{equation}
p(T) = A \frac{T^2}{\Lambda} ~~~,
\end{equation}
where $A$ is a normalization constant, which we can write as 
\begin{equation}
\begin{split}
\fmm &~ = \frac{M_{\rm cooling}}{\mcool} = \int{p(T)dT} = \\
     &~ = A \int_{\teq}^{T{\rm warm}}{\frac{T^2}{\Lambda} dT} \equiv A \times I_T(\teq,T_{\rm warm}) ~~~,
\end{split}
\end{equation}
The value of the upper integration boundary is motivated by our goal to model the gas that is actively cooling. Then the temperature and density PDFs are given by
\begin{equation}
p(T) = \frac{\fmm}{I_T(\teq,T_{\rm warm})} \frac{T^2}{\Lambda} ~~,~~ g(n) = \frac{\fmm}{I_n(\teq,T_{\rm warm})} \frac{1}{\Lambda n^4} ~~~,
\end{equation}
where $I_n$ is the integral value for the density distribution. We note that the full PDFs extend to higher temperatures (and lower densities), including the hot ambient gas in the CGM at $T_{\rm hot}$. Integration over the full temperature range gives the hot to cool CGM mass
\begin{equation}\label{eq_app:hot_prob}
\frac{M_{\rm hot}}{\mcool} = \fmm \frac{I_T(\teq,T_{\rm hot})}{I_T(\teq,T_{\rm warm})} = 
\fmm \frac{\int_{\teq}^{T{\rm hot}}{T^2/\Lambda dT}}{\int_{\teq}^{T{\rm warm}}{T^2/\Lambda dT}} ~~~.
\end{equation}
Given the gas metallicity and cooling function, and the model $n_0$ and $\fmm$, we can calculate the hot to cool gas mass ratios for different values of $T_{\rm warm}$ and $T_{\rm hot}$. In this work we adopt $T_{\rm warm} = 2 \times 10^5$~K, in the middle of the cooling function peak for metal-enriched gas ($Z>0.1$), and $T_{\rm hot} = 10^6$~K, motivated by the approximate virial temperature for a MW-mass halo at low-$z$. For these temperatures and $\fmm = 0.6$ we get $M_{\rm hot}/\mcool = 45$ at $n_0 = 10^{-3}~\cmv$, and $M_{\rm hot}/\mcool \approx 40-50$ at $n_0 = 10^{-4}-10^{-2}~\cmv$. These ratios are consistent with $\mcool \approx 0.1-1.0 \times 10^{10}~\msun$ and $M_{\rm hot} \approx 0.3-1.0 \times 10^{11}~\msun$.

We note that high values of $T_{\rm hot}$ produce large ratios of $M_{\rm hot}/\mcool$, which imply either small $\mcool$ that cannot reproduce the observed columns of low ions, or $M_{\rm hot}$ that exceed the galactic baryonic budget. For example, for $\fmm=0.6$, $T_{\rm warm} = 10^5$~K and $T_{\rm hot} = 2 \times 10^6$~K give $M_{\rm hot}/\mcool = 300$. Mass ratios that are consistent with observational estimates thus provide another constraint on the PDF temperature limits for this model.

\vspace{0.5cm}
We note that the normalization of the mixing layers PDF can also be done at an intermediate temperature of the mixing gas, similar to how it is performed for the cooling gas scenario. However, due to the flat shape of the mixing layers PDF, leading to a weak dependence of the normalization on the integral boundary temperature, this introduces only a small difference.

\restartappendixnumbering
\section{Metallicity Variation}
\label{app_met_var}

In this work we assume a nominal metallicity of $Z=0.3$ solar, following the median density of the absorbers inferred by \citetalias{Prochaska17}. The assumed metallicity affects the gas ionization state through the gas equilibrium temperature in PIE (see \ref{sec:methods} and Figure~\ref{fig:temp_eq})

Figure~\ref{fig:metallicity} shows the effect of metallicity variation on some of the ion fractions we consider in this work -- \NII\ and \NIII\ (left panels), \CIII\ and \CIV\ (middle), and \SiIII\ and \SiIV (right) -- as function of the gas density. The black and red curves show the low and high ion in each panel, respectively. In the top panels we plot the fractions at the PI equilibrium temperature for the nominal metallicity assumed in this work, $Z=0.3$ solar, using solid curves. The dashed and dotted curves show the fractions for $Z=0.1$ and $Z=1.0$, respectively. The bottom panels show the ratios of the ion fractions with these metallicities to the ion fractions at the nominal $Z$.

Most of the ions presented here show only a weak sensitivity to the gas metallicity and temperature. For the gas densities relevant for this work, $3 \times 10^{-4} < \nh/\cmv < 10^{-2}$ (see \citetalias{Prochaska17} and Table~\ref{tab:fit_results} here), highlighted in grey in Figure~\ref{fig:metallicity}, a factor of $3$ variation in metallicity translates to an effect of $<25\%$ in ion fraction, compared to $Z=0.3$. For the carbon ions, this is true for the entire density range presented here, and for nitrogen the ion fractions only deviate more significantly at densities below $\simeq 10^{-4}~\cmv$. The silicon ions show a slightly stronger sensitivity to the gas metallicity, with deviations of up to $\approx 50\%$ in the shaded density range for $Z=1$ solar (dotted curve), corresponding to lower $\teq$ and lower ionization. \SiIII\ has a low fraction at higher densities (black curve), with more silicon in \SiII\, and \SiIV\ - a higher fraction at lower densities, with less silicon in $\SiV$. For the full density range plotted, the ion fraction of these two ions can be lower/higher than those of the nominal metallicity by up to $70\%$.

When fitting in \S\ref{subsec:fit_method}, we increase most errors on $\log$ column to $\sigma_N=0.1$, or $60\%$ uncertainty.  This is partly  motivated by this difference.  We could improve our theoretical model by accounting for the $\teq$ dependence on metallicity, but we note that other heating sources like turbulence or cosmic rays could result in somewhat higher temperatures that we expect could be on par with the temperatures from changing the metallicity.  We note it is difficult at these densities to get temperatures below 8000K or above 20000K for equilibrium gas because of the strong sensitivity of \HI\ cooling to temperature.

We note that these are the ion fractions at the gas equilibrium temperature, without density or temperature distributions. Once fluctuations are included, the effect of metallicity on the effective fractions, averaged over a range of densities and temperatures independent of $\teq$ (see \S\ref{subsec:distributions}) will be smaller than shown here.

\begin{figure*} \label{fig:metallicity}
{\centering{\includegraphics[width=0.98\textwidth]{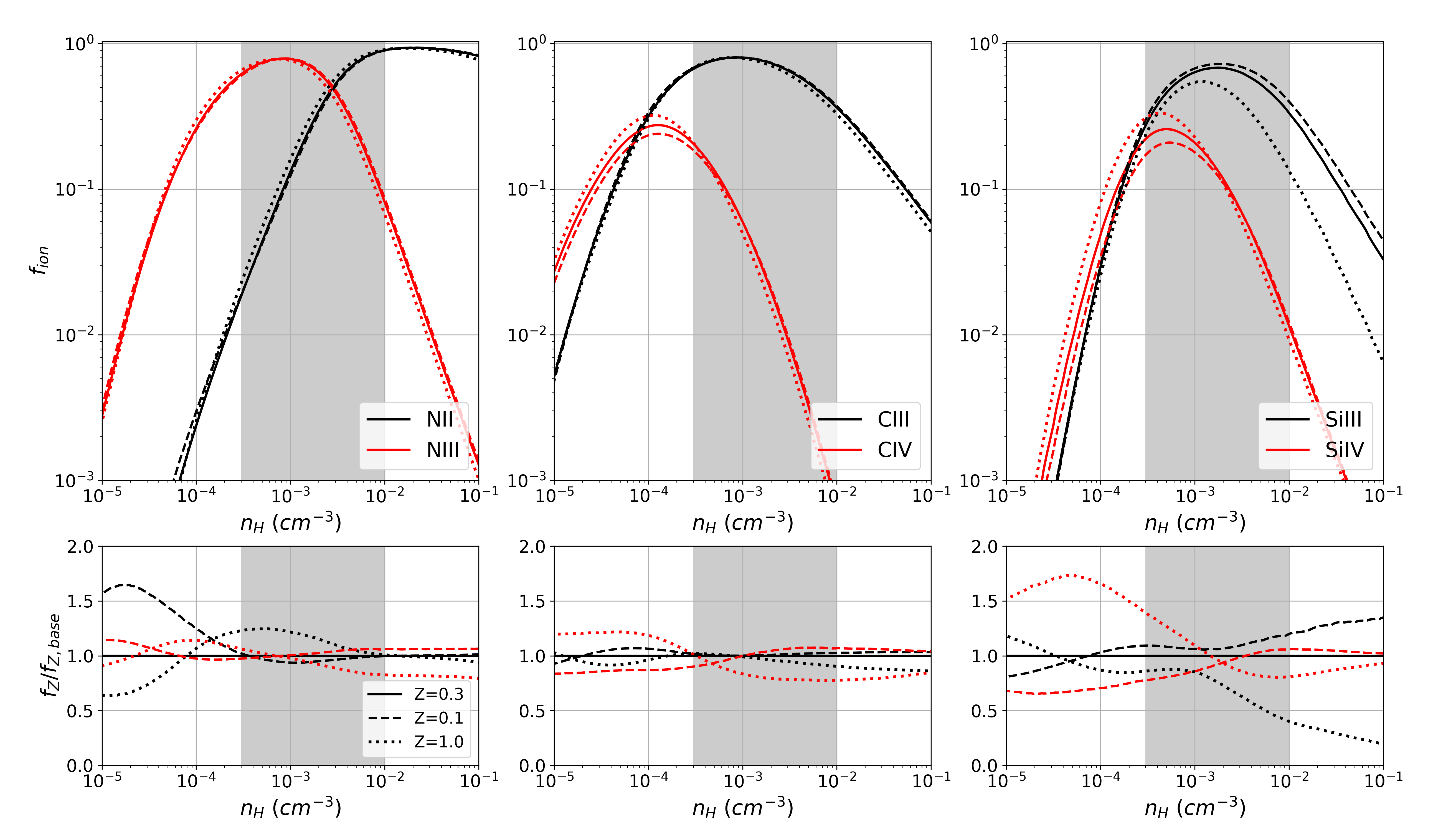}}}
\caption{The effect of metallicity on the ionization state at PIE. The gas metallicity affects the ion fractions through the equilibrium temperature (see Figure~\ref{fig:temp_eq}). The top panels show the ion fractions for $Z =0.3$, the nominal metallicity assumed for the CGM in this work (solid curves), and for $Z=0.1$ and $Z=1.0$ (dashed and dotted, respectively). The bottom panels show the ratios of the ion fractions at lower and higher metallicities to that at $Z=0.3$. For the density range of our absorbers (shaded grey, see Table~\ref{tab:fit_results}), a factor 3 metallicity variation has a small effect on most ions, with a change of $<25\%$ in ion fraction compared to the nominal. Two exceptions are the silicon ions which exhibit a more significant deviation from the nominal values, up to $50\%-70\%$, at high metallicities (corresponding to low $\teq$). \SiIII\ has a lower fraction at higher densities and \SiIV\ - a higher fraction at lower densities.}
\end{figure*}

\restartappendixnumbering
\section{\HeII\ Self-Shielding}
\label{app_HeII}

As described in \S\ref{sec:fit_testing}, in this work we select our sample so that we can neglect the effects of \HI\ self-shielding, by choosing objects with $N_{\rm HI} < 10^{17.5}~\cmc$. However, self-shielding by \HeII, which absorbs radiation at $>4$~Ry, may still be important. We calculate the ion fractions for the optically thin and fully self-shielded cases and perform our fitting to observations for both cases to examine whether the data prefers one of the scenarios (see Table~\ref{tab:fit_results}). We also calculate the \HI\ critical column for \HeII\ self-shielding as a function of $\nh$ and examine whether our analysis and assumptions are consistent given the measured \HI\ columns and our inferred $\nh$ values (see Figure~\ref{fig:nhi_crit}). In this section we examine how self-shielding affects the individual fractions of selected ions.

\begin{figure*} \label{fig:selfshielding}
{\centering{\includegraphics[width=0.98\textwidth]{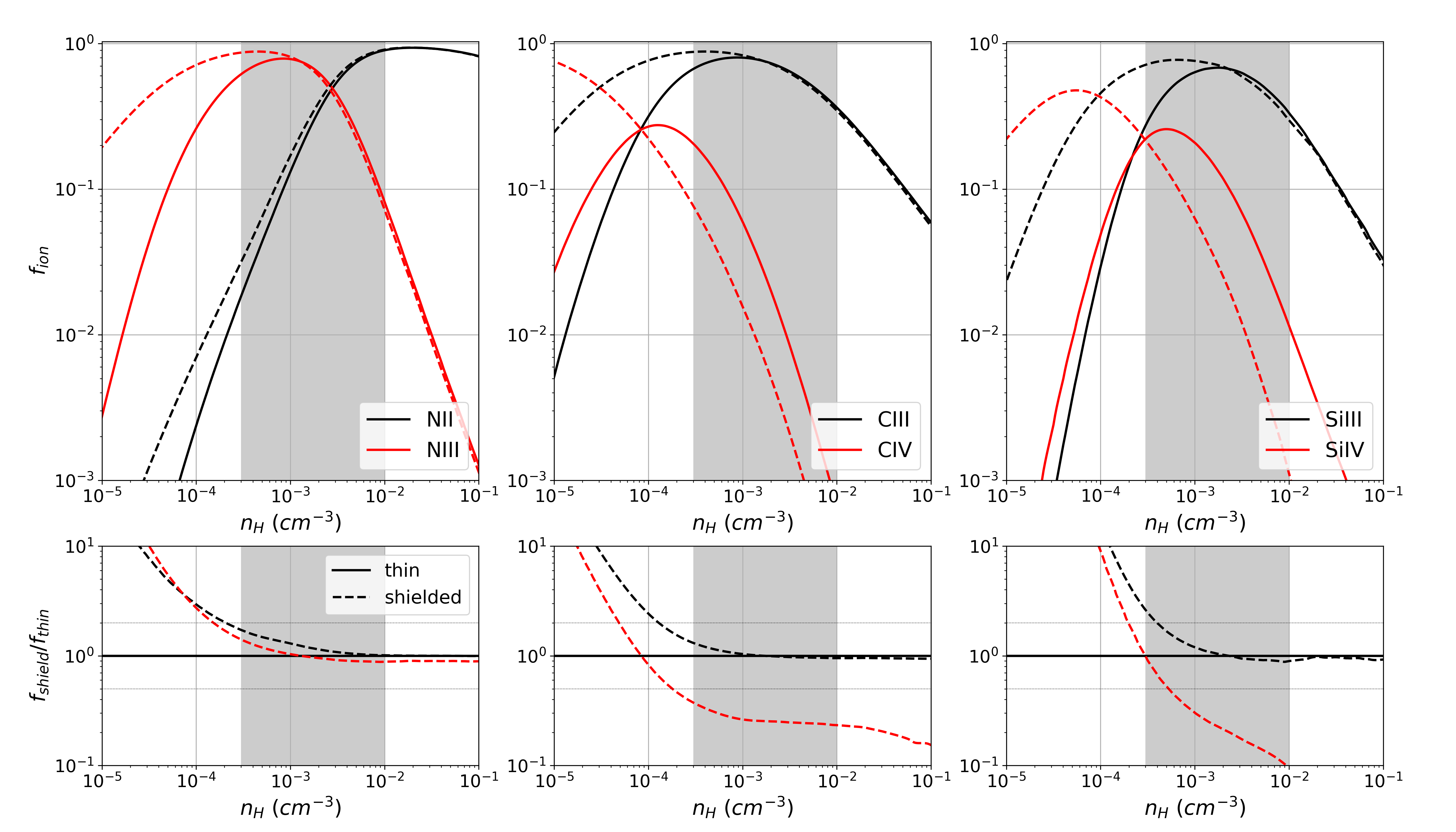}}}
\caption{Effect of \HeII\ self-shielding on prominent ions. The solid curves show the ion fractions assuming that the \HeII\ column density is sufficiently small that the $>4$~Ry MGRF is not absorbed, whereas the dashed curves are for the other extreme, in which these energies are fully absorbed. The grey shaded region highlights the densities inferred from the COS-Halos observations (see Table~\ref{tab:fit_results}). The \HI\ columns of low-$z$ CGM clouds suggest that the $4$~Ry background is likely to be significantly absorbed only for a few of the objects in our sample (see Figure~\ref{fig:nhi_crit}).}
\end{figure*}

Our results are plotted in Figure~\ref{fig:selfshielding}. We examine the same set of ions shown in Figure~\ref{fig:metallicity} - \NII\ and \NIII\ (left panel), \CIII\ and \CIV\ (middle), and \SiIII\ and \SiIV\ (right). The top panels show the ion fractions, and the solid curves are for the optically thin case, in which the \HeII\ column density is sufficiently small that the $>4$~Ry MGRF is not absorbed, whereas the dashed curves assume the other extreme, that these energies are fully absorbed.  As the highest ions we study have ionization potentials that are not much greater than $4~$Ry, this approximation likely encapsulates the effect of self shielding once the $4$Ry optical depth is a factor of a few greater than unity. The bottom panels show the ratio of ion fractions in fully self-shielded gas over the optically-thin values.

First, we see that the low and intermediate ions (\NII, \NIII, \CIII, and \SiIII) are only slightly affected, if at all, at densities above $10^{-3}~\cmv$~\footnote{As we discuss in \S\ref{subsec:redshift}, we perform our calculations at $z=0.2$ and due to the evolution of the MGRF ionizing flux with redshift, the density is given by $\nh = 10^{-3} [(1+z)/1.2]^{\alpha}~\cmv$, with $\alpha\approx 4.5$ at $z<0.5$.}. At lower densities they can be higher, as in the self-shielded case gas is less ionized and a larger fraction of a given element is in these lower ionization states. At densities in the range relevant for this work (shaded grey), the fractions of these ions are within a factor of $\lesssim 2$ from their values in optically thin gas (bottom panels).

Second, the fractions of higher ions (\CIV\ and \SiIV) can be reduced significantly in self-shielded gas and at at densities above $1-3 \times 10^{-4}~\cmv$, and in the range of densities relevant for this work they are lower by factors of 3-10.

\label{lastpage}
\end{document}